\pdfoutput=1

\documentclass{gMOS2e_hacked_by_TU}

\usepackage{amsmath} 

\usepackage{subfig}

\usepackage{graphicx}

\usepackage{url}

\newcommand{\D}{DL\_MONTE\ }
\newcommand{\Dns}{DL\_MONTE}
\newcommand{\DD}{DL\_MONTE\ }
\newcommand{\DP}{DL\_POLY}

\begin{document}


\title{DL\_MONTE: A multipurpose code for Monte Carlo simulation}

\author{ 
 \name{A.~V.~Brukhno\textsuperscript{a$\dagger$}\thanks{$^\dagger$ E-mail: andrey.brukhno@stfc.ac.uk}, J.~Grant\textsuperscript{cd}, T.~L.~Underwood\textsuperscript{b$\ddagger$}\thanks{$^\ddagger$ E-mail: t.l.underwood@bath.ac.uk}, 
       K.~Stratford\textsuperscript{e}, S.~C.~Parker\textsuperscript{c}, J.~A.~Purton\textsuperscript{a}, 
       and N.~B.~Wilding\textsuperscript{b}
      }
 \affil{
       \textsuperscript{a}Scientific Computing Department, STFC, Daresbury Laboratory, Keckwick Lane, Warrington WA4 4AD, UK;
       \textsuperscript{b}Department of Physics, University of Bath, Bath BA2 7AY, United Kingdom;
       \textsuperscript{c}Department of Chemistry, University of Bath, Bath BA2 7AY, United Kingdom;
       \textsuperscript{d}Computing Services, University of Bath, Bath BA2 7AY, United Kingdom;
       \textsuperscript{e}EPCC, University of Edinburgh, EH9 3FD, Edinburgh, United Kingdom
      }
}

\maketitle

\begin{abstract}
\D is an open source, general-purpose software package for performing Monte Carlo simulations.
It includes a wide variety of force fields and MC techniques, and thus is applicable to a broad range of 
problems in molecular simulation. Here we provide an overview of \Dns, focusing on key features 
recently added to the package. These include the ability to treat systems confined to a planar 
pore (i.e. `slit' or `slab' boundary conditions); the lattice-switch Monte Carlo (LSMC) method for evaluating precise 
free energy differences between competing polymorphs; various commonly-used methods for evaluating 
free energy profiles along transition pathways (including umbrella sampling, Wang-Landau and 
transition matrix); and a supplementary Python toolkit for simulation management and application
of the histogram reweighting analysis method. We provide two `real world' examples to elucidate the use 
of these methods in \Dns. In particular, we apply umbrella sampling to calculate the free energy profile associated 
with the translocation of a lipid through a bilayer. Moreover we employ LSMC to examine
the thermodynamic stability of two plastic crystal phases of water at high pressure.
Beyond this, we provide instructions on how to access \Dns, and point to additional information valuable to 
existing and prospective users.
\end{abstract}


\section{Introduction}\label{sec:intro}

Computational modelling is often cited as the third pillar of science along with experiment and theory. More specifically, 
molecular simulation provides powerful, detailed insights and helps our understanding of condensed matter and materials on 
atomistic and nano scales~\cite{Allen_Tildesley_87,Frenkel_Smit}. Moreover its predictive capacity is utilised in industry to 
guide the development of new and more effective products, cutting development costs, reducing time to market, and improving
manufacturing efficiency~\cite{Meunier_2012}.

The two workhorse methods in molecular simulation are Monte Carlo (MC) and molecular dynamics (MD). Both methods entail sampling 
configurations from a specified thermodynamic ensemble, e.g. the canonical ($NVT$) ensemble or the isobaric-isothermal ($NPT$) 
ensemble. In some situations, MD is superior to MC because it employs `realistic' Newtonian dynamics, and hence can be used to 
study kinetic processes and determine time-dependent quantities such as molecular vibrations and diffusion constants. 
Moreover, MD also parallelises efficiently, making it especially suitable for treating very large systems.
However, there are many situations where MC is preferable: for instance, in simulations of `open' or highly non-uniform (inhomogeneous) systems, where particles can enter and leave the system, or tend to form aggregates. The ability to deal with such systems is important when studying the 
adsorption of gases at surfaces and within porous materials~\cite{Knight_1986,Fan_2011,Gatica_2008}, for fluids, phase 
transitions~\cite{Wilding_2001,Brukhno_2009,Brukhno_2011}, and for multi-component mixtures~\cite{Cracknel_1995,Lima_2012}.

Owing to its ability to exploit `unphysical' particle dynamics or creative thermodynamic ensembles, MC is typically more powerful and versatile than MD in addressing the sampling issues which arise from rough (free) energy landscapes, entropic bottlenecks and extended correlation times. 
Pertinent examples here include Gibbs ensemble MC ~\cite{Panagiotopoulos_1987} for studying phase coexistence; lattice switch 
MC~\cite{Bruce_1997} for computing precise free energy differences between competing polymorphs; and replica exchange (also known
as parallel tempering)~\cite{Swendsen_1986,Earl_2005} for accelerating sampling in `glassy' energy landscapes.  
Noteworthy too are the variety of MC methods for calculating the free energy with respect to some transition pathway that has been parameterised
in terms of an order parameter or reaction coordinate. Commonly used approaches here include umbrella sampling~\cite{Torrie_Valleau_1974,Torrie_Valleau_1977}, adaptive 
umbrella sampling~\cite{Mezei_1987,Mezei_1989}, expanded or {\it generalised} 
ensembles~\cite{Lyubartsev_1992,Broukhno_2000,Okamoto_2000,Iba_2001}, entropic sampling~\cite{Lee_1993} (enhanced with Wang-Landau 
bias feedback scheme~\cite{Wang_2001}), and the transition-matrix method~\cite{Smith_1995,Fitzgerald_1999}.

The past few decades has seen the emergence of a variety of sophisticated `general-purpose' MD simulation 
programs -- DL\_POLY~\cite{dlpoly1, dlpoly2} among them. These MD packages have facilitated the advance of MD into many 
fields, rendering it an invaluable tool for tackling `real world' scientific problems. Unfortunately the same
cannot yet be said of MC simulation. Implementations of MC methods have traditionally been limited to in-house 
codes, tailored to a specific problem. General-purpose MC programs that provide access to a wide range of techniques -- including advanced techniques for complex systems --  have been 
lacking. As far as we are aware there are only a handful of general-purpose MC programs available to the community 
\cite{casandra,music,raspa,towhee,Purton_2013}, and of these, only Casandra~\cite{casandra} and \D~\cite{Purton_2013} are under 
active development.
However, the development of such programs is essential if the unique capabilities of MC are to be fully exploited by the scientific 
community. 

The \D project was initiated under the auspices of EPSRC~\cite{epsrc} and CCP5~\cite{ccp5}, with the aim of providing MC software
which:
\begin{enumerate}
\item includes a wide variety of force fields, as well as `standard' MC functionality (for example the ability to simulate atoms 
and molecules in 
the $NVT$, $NPT$ and $\mu VT$ ensembles), making it suitable for use in broad academic research;
\item includes various state-of-the-art MC methods, facilitating the uptake of these methods by the scientific community;
\item is open source, accessible, and well documented;
\item is cross-compatible with DL\_POLY as much as possible, thus acting as a complementary MC alternative to DL\_POLY.
\end{enumerate}
In this work we present \D (version 2), with particular emphasis on the extensive functionality which has been added to the 
program since the release of version~1 in 2013~\cite{Purton_2013}. The most important additions include the lattice-switch MC 
method; and the widely-used methods for calculating free energy profiles -- hereafter collectively referred to as 
\emph{free energy difference} (FED) methods -- which can be used in conjunction with replica-exchange parallel tempering.
Another new feature is the ability to treat systems confined to a planar pore, i.e. in `slit' or `slab' geometry. Regarding this, 
numerous types of wall-particle potential are provided, including hard or 
soft walls, and walls bearing surface charge density. Long-range electrostatics are also supported in both conventional 
(i.e. periodic 3D) and slit geometries. 

Finally, \D has been equipped with a Python-based simulation management and analysis toolkit. This includes 
both programming and in-browser iPython interfaces for execution of the code and manipulation of simulation input and output. The 
toolkit also implements the powerful multi-histogram reweighting analysis method~\cite{Ferrenberg_1988} as an extensible Python API
class, as well as a self-contained weighted histogram analysis method (WHAM)~\cite{Ferrenberg_1989} utility that is ready to 
apply directly to FED output data. 

The paper is organised as follows. We begin by providing a brief overview of the principal  functionality of \D in Section~\ref{sec:functionality} (technical details regarding workflows for parallel simulation  and performance optimisation are deferred to Appendix~\ref{sec:performance}). Next, in Section~\ref{sec:access}, 
we provide instructions on how to access the software, and also point to sources of additional information valuable for existing and 
potential users. In Section~\ref{sec:toolkit} we introduce the Python toolkit and demonstrate its functionality and usage. 
Section~\ref{sec:fed} describes in some detail the theory which underpins the key FED implementations, including lattice-switch MC, and 
describes how we have validated the functionality against known results. Then in Section~\ref{sec:applications} we present 
two `real-world' example applications.  The first demonstrates the capability to treat complex molecular systems by employing umbrella sampling to calculate the free energy profile associated with the translocation of a lipid molecule across a lipid bilayer. The second example deploys the lattice-switch MC capability to study the relative stability of two plastic crystal 
phases of a water model at high pressure. Section~\ref{sec:summary} provides a summary of the paper.

\section{Overview of functionality}\label{sec:functionality}

In this section we outline the principal functionality  of \Dns. Further details including descriptions of various simulation workflows and input/output data files can be found in the user manual and hands-on tutorials, access to which is described in Section~\ref{sec:access}. Note that, with the exception of the FED methodology which is elaborated on in Section~\ref{sec:fed}, we shall not cover the well-known general theory underpinning standard Monte Carlo (Metropolis) algorithms and their implementation in \Dns. Uninitiated readers who are interested in learning more about both MC and MD techniques are referred to the many comprehensive textbooks on molecular simulation, see e.g.~\cite{Allen_Tildesley_87,Frenkel_Smit}.

\subsection{Force fields and particle dynamics}
In \D the system is abstracted into `atoms' and `molecules': atoms are treated as point-like particles, and molecules are collections of atoms which can be moved collectively. This, along with a versatile selection of potential forms which can be combined into different force fields, allows for simulation of a wide range of systems -- including fluids, colloids, inorganic solids, semiconductors, metals, and biomolecules. These include systems comprised of combinations of `free' unconnected atoms (so-called `atomic field'), and molecules possessing structure: rigid and flexible ones.

For the purposes of Monte Carlo simulation, an {\it MC force field} is, by definition, a collection of energy terms contributing to the Hamiltonian of a given system, owing to the fact that the actual forces acting between `atoms' are, generally, irrelevant and not used in MC. That said, adopting the commonly used terminology, the force field definitions in \D largely follow the DL\_POLY conventions. Apart from the basic properties of `atoms' (e.g. type, mass, charge), their pairwise and possibly multi-body, so-called {\it non-bonded}, interactions, the input also includes definitions of the topology (i.e. internal structure) of all the distinct molecular species present. Thus, part of the force field determines intra-molecular, so-called {\it bonded}, interactions, such as chemical bonds (or, generally, connectivity between more abstract `monomeric' units), bending, dihedral (torsion) and inversion angles within a molecule. 

In general, the force field contributions to the Hamiltonian can be categorised by a few major interaction types:
\begin{itemize}
\item long-ranged pairwise electrostatic interactions acting between charged atoms (if present);
\item short-ranged pairwise van der Waals (VDW) interactions acting between non-bonded atoms which can either belong to different molecules or sit on the same molecule;
\item three-body interactions: non-bonded and/or bonded, e.g. bending angles in molecules (if present);
\item bonded four-body interactions: torsion and inversion  angles in molecules (if present);
\item many-body non-bonded interactions: Tersoff and metal potentials (if present);
\item interaction of atoms with an external field (if present).
\end{itemize}
A number of widely used functional forms are supported for the two-, three- and four-body interactions, and up to five additional pairwise (VDW) interactions can be defined in analytical form by the user (see the \D user manual). Within a given force field, the listed interaction types can be combined. However, to avoid ambiguity, only one specific interaction from each category can be applied to a particular set of atoms at the same time. We also note that pairwise exclusion rules can be specified for VDW and Coulomb {\it intramolecular} interactions between the following pairs pertaining to the structural elements within a molecule: 1-2 (bonds), 1-3 (bending angles), and 1-4 (dihedral and inversion angles). The exclusion list is aimed to mimic exclusions in the known conventional force fields (e.g. CHARMM, Martini etc.).

For generating new configurations \D implements six standard MC moves: (1) atom translation, (2) molecule translation, (3) molecule rotation, (4) atom insertion/deletion and (5) molecule insertion/deletion, (6) pairwise swapping of atoms or molecules, which can be used in combination, of course. This set of generic MC moves proved to be sufficient for the simulation scenaria considered in this paper, whereas more sophisticated moves are planned for addition in the future, e.g. pivot moves, configuration bias~\cite{Frenkel_Smit,Siepmann_1992} and geometric cluster algorithm~\cite{Dress_1995,Liu_2005}.


\subsection{Boundary conditions}
Along with conventional 3D periodic boundary conditions, the program also supports the planar pore (or `slit') geometry with quasi-2D boundary conditions, in which the system is periodic in the $X$ and $Y$ directions, but confined in the $Z$ direction. The slit constraint is enhanced with an extensive set of external potentials, including most of the available short-range (VdW) types re-defined as particle-surface interactions. 

Two approaches are available for including the long-range corrections to electrostatic interactions in quasi-2D slit geometry. In the first case, with true non-periodic $Z$-dimension, a computationally inexpensive approach is to employ a mean-field approximation (MFA) for the Coulomb interactions between the charges in the primary cell and the external charge density outside of the simulation cell (which is set equal to the charge distribution within the cell)~\cite{Torrie_1980,Valleau_1991,Broukhno_2002,Brukhno_2011}. This self-consistent MFA scheme works best for unstructured fluids with high dielectric permittivity, e.g. solvent-free CG models. The second approach, which is often used in MD simulations of confined solutions, is to utilise a so-called `slab' arrangement within a normal fully periodic simulation cell~\cite{Yeh_1999,Tieleman_2002,Bostick_2003}. In this case, the $Z$-dimension of the simulation cell is extended and filled with vacuum beyond the actual slit confinement. The regular (3D) Ewald summation method can then be employed, provided the Coulomb interactions vanish in the $Z$ direction within the extended vacuum portion. 

\subsection{Thermodynamic ensembles}

As well as the canonical ($NVT$) ensemble, simulations can be performed in other thermodynamic ensembles:
\begin{itemize}
\item isobaric-isothermal ($NPT$) and isotension-isothermal ($NP_{xy}T$), where MC moves 
attempting variations in the volume of the system are applied; 
\item grand canonical ($\mu VT$), where atoms or molecules are added and removed from the system while maintaining
the system at a fixed chemical potential $\mu$;
\item semi-grand canonical ensemble, where the total number of atoms/molecules is fixed, but the concentrations of 
species can change via identity swaps, i.e. pairwise `mutations' of atoms or molecules while the difference between the chemical potentials of the two species involved is kept constant.
\end{itemize}
Note that the grand and semi-grand canonical ensembles are `open' ensembles -- particles
are effectively exchanged with (virtual) external reservoirs in the course of a simulation. As mentioned in Section~\ref{sec:intro}, open ensembles are often more efficient in simulation of chemical equilibria and analysis of chemical composition (with relatively small moieties) than closed ensembles. 

Also mentioned in Section~\ref{sec:intro} was the fact that \D implements a number of advanced methods which go beyond the traditional thermodynamic ensembles, namely:
\begin{itemize}
\item Gibbs ensemble MC~\cite{Panagiotopoulos_1987}, in which two coexisting phases are simulated simultaneously in a single simulation, without the requirement of creating an interface between them. This method is commonly used to study vapour-liquid and liquid-liquid equilibria.
\item Replica exchange parallel tempering,~\cite{Swendsen_1986,Earl_2005} in which multiple replicas of the system are simulated simultaneously at different temperatures. The characteristic feature of this method is that the replicas are coupled: MC moves which `swap' configurations belonging to different temperatures are attempted periodically. The end result is improved sampling efficiency in the low-temperature copies 
of the system if the energy landscape has many competing local minima.
\item Various FED methods which collectively can be regarded as generalised ensemble methods:
harmonic umbrella sampling~\cite{Torrie_Valleau_1974,Torrie_Valleau_1977}, expanded ensemble~\cite{Lyubartsev_1992,Broukhno_2000}, Wang-Landau scheme for on-the-fly bias optimisation~\cite{Wang_2001}, transition-matrix~\cite{Smith_1995,Fitzgerald_1999} and lattice-switch MC~\cite{Bruce_1997}. 
\end{itemize}
We elaborate on \Dns's capability in regard to FED methods, as well as the theory which underpins these methods, later in Section~\ref{sec:fed}. Lattice-switch MC, a method for evaluating the free energy difference between two given solid phases to high precision, which draws heavily on the FED methods, is also described in Section~\ref{sec:fed}.

\subsection{Performance and optimisation}
\D has a number of features and controls for tweaking the efficiency of a simulation. These include: tunable neighbour lists (auto-updated or user-tailored), automatic rejection of MC moves resulting in particles found within a pre-defined distance from each other, and two modes of loop parallelisation: atom-wise or molecule-wise. 
As is common for simulation packages, \D can be compiled and run in a high-performance computing (HPC) environment (e.g. on Beowulf clusters) with the use of MPI libraries (to be pre-installed separately), which enables its internal parallelisation of the most expensive calculations: the energy updates and the Ewald summation for long-range electrostatics. More detail on the aspects of optimisation and parallelisation can be found in Appendix~\ref{sec:performance}.


\subsection{Other features}
\D also has a number of features which facilitate its general usability. For instance, as well as the Python toolkit, which is discussed in detail in Section~\ref{sec:toolkit}, \D has the ability to store configurations and trajectories in conventional, commonly-used formats, such as DL\_POLY (2 and 4) text and DCD (CHARMM/NAMD) binary formats which are compatible with third-party visualisation (VMD~\cite{vmd}) and analysis (Wordom~\cite{wordom}) packages. Moreover \D has the ability to convert between these formats and the native \D format. 
This assists greatly with visualising trajectories obtained from simulations, as well as data analysis.

\section{Accessing and using \Dns}\label{sec:access}
The \D homepage can be found at \url{http://www.ccp5.ac.uk/DL_MONTE}. This is the primary source of information on 
the program, including information regarding upcoming releases, training events, etc. However the program itself 
is hosted on CCPForge at \url{http://ccpforge.cse.rl.ac.uk/gf/project/dlmonte2/}. 
To access this, one must first register an account with CCPForge, 
and then request to join the project \verb|DL_MONTE-2|. 
(Note that the project name is \verb|DL_MONTE-2|, \emph{not} \verb|DL_MONTE|, which pertains to the version 1
of the program, and is no longer active). 
Once the user's request to join the \verb|DL_MONTE-2| project is approved (usually within 24 hours), 
they can then download a release of \Dns.

\subsection{Software dependencies}

\D is self-contained in that it does not crucially depend on any third-party libraries or modules.
To compile the serial version of the program all that is required is a Fortran 95 compiler. However to compile the 
parallel version a standard MPI library is required.

\subsection{Licence}
\D is free software and open source, released under a BSD licence.

\subsection{Usage and user support}

Included with a release of \D is a user manual, which details its functionality, usage, and how to compile it. 
The user manual for the latest release is always publicly visible on CCPForge (i.e. the \verb|DL_MONTE-2| project on CCPForge, \url{http://ccpforge.cse.rl.ac.uk/gf/project/dlmonte2}),
so to give prospective users insight into the program before downloading it. While the manual is an invaluable resource for users, there are also a set of tutorials which provide a pedagogical introduction to \D and MC methodology. These can also be obtained from CCPForge.

User support is provided through a forum on CCPForge, where users can 
flag bugs, provide feedback, and ask developers for assistance with using \Dns.

\section{Python toolkit}\label{sec:toolkit}

Solving a given problem using molecular simulation is rarely as simple as performing
a single simulation and analysing its output. Typically complex workflows must be employed
which involve cycles of running one or more simulations, analysing their output, 
and then using the results of this analysis to inform input parameters for further simulations. 
Software which helps manage simulation workflows is therefore of great interest.
Such software is usually provided as a separate set of helper utilities, or `toolkit', 
written in a scripting language. In this respect Python is very attractive, since it provides 
users with the means to adapt and manage their particular workflows in a flexible manner.

Motivated by this, we have developed a Python toolkit for managing workflows
involving \Dns. There are two key facets to the toolkit. Firstly, it provides
Python interface to \Dns, enabling users to execute simulations,
as well as manipulate the input to, and output from, simulations from
within a Python environment. Secondly, the toolkit provides an implementation of the 
\emph{histogram reweighting} analysis technique~\cite{Ferrenberg_1988}. 
This technique takes data obtained at a certain set of thermodynamic parameters 
(e.g. temperature, pressure) and uses it to make predictions about the properties of the
same system at a different set of parameters.
For example one could use histogram reweighting to deduce, from 
data obtained from a simulation conducted at 300K, the properties of the same system
at 310K, \emph{without the need to perform another simulation at 310K}. Histogram
reweighting is useful because it allows one to `make the most' of the simulation data
one already has, perhaps reducing the computational resources required to solve the 
problem at hand.

In the rest of this section we provide a brief description of the toolkit. We begin by describing
how the toolkit can be used to execute \D and interface with input and output files.
We then describe the histogram reweighting aspect of the toolkit in more detail, presenting
some results obtained using the toolkit which elucidate the method.

\subsection{Python interface to \Dns}\label{sec:toolkit_interface}
The toolkit provides Python classes which represent \D input and output files 
in the form of a structured object.
Moreover, there is a class which represents all input files collectively 
as a single Python object, and similarly for output files. These classes, and associated convenience 
functions, facilitate manipulation of input, output, and simulation control parameters,
and extraction of pertinent output data from within a Python environment. In addition 
to these data-structure classes, there is also a global class that unites those mentioned above,
and allows \D to be executed with the input taken from a specified directory.
These all provide a complete framework for creating semi-automated, customizable workflows involving \Dns.

An example Python script that demonstrates this aspect of the toolkit is given in Appendix~\ref{sec:pythonscript}. 
The script imports input parameters from a 
directory containing input files; 
uses these parameters as a template, and runs a set of simulations at various temperatures; and finally
analyses the data from the simulations to deduce the mean energy of the system vs.
temperature, printing the energy vs. temperature to standard output.

\subsection{Application of the toolkit: Histogram reweighting}
As mentioned in Section~\ref{sec:intro}, an MC or MD simulation samples configurations from the 
probability distribution corresponding to the thermodynamic ensemble under consideration. For a set
of $n$ uncorrelated configurations obtained from an MC/MD simulation, one can calculate the expected
value for some observable $O$ for the underlying ensemble via 
\begin{equation}
\langle O\rangle=\frac{1}{n}\sum_{i=1}^nO_i,
\end{equation}
where $O_i$ is the value of the observable for the $i$th configuration. The above equation can be
recast as follows:
\begin{equation}\label{expected_O}
\langle O\rangle=\sum_{i=1}^nw_iO_i \Bigg/ \sum_{i=1}^nw_i,
\end{equation}
where $w_i$ is the \emph{weight} applied to configuration $i$ in the evaluation of 
$\langle O\rangle$, and here all configurations have equal weight, e.g. $w_i=1$ for all $i$.

Typically the probability distribution for the considered thermodynamic ensemble is 
known. For example, in the canonical ($NVT$) ensemble the probability associated with 
configuration $i$ is 
\begin{equation}
p_i\propto\exp(-\beta E_i),
\end{equation}
where $E_i$ is the energy of $i$ and 
$\beta\equiv 1/(k_BT)$ is the inverse temperature. We can exploit this to take data 
obtained from a simulation conducted at one value of a thermodynamic parameter 
(e.g. temperature, inverse temperature, pressure, or chemical potential)
and use it to calculate the expected value of $O$ at a \emph{different} thermodynamic 
parameter. This is achieved by altering the weights $w_i$ in Eq.~\ref{expected_O} such
that they pertain to the `new' parameter. For example, if our simulation were performed
in the canonical ensemble at inverse temperature $\beta$, and we were interested in using
our simulation data to calculate $\langle O\rangle$ at a different inverse temperature $\beta'$,
then we could exploit the fact that the probability $p_i$ associated with a configuration $i$ at 
$\beta$ is related to the probability $p_i'$ of the configuration at $\beta'$ via
\begin{equation}
p_i'\propto p_i\exp\Bigl[-(\beta'-\beta)E_i\Bigr].
\end{equation}
Hence applying Eq.~\ref{expected_O} with
\begin{equation}\label{reweight_w}
w_i=\exp\Bigl[-(\beta'-\beta)E_i\Bigr]
\end{equation}
will yield $\langle O\rangle$ corresponding to inverse temperature $\beta'$.
One can say that the data at $\beta$ has been \emph{reweighted} to a new inverse temperature $\beta'$ -- for 
the purposes of evaluating $\langle O\rangle$. 
Reweighting can also be applied to other thermodynamic parameters in other ensembles.
Moreover reweighting in more than one thermodynamic parameter, and reweighting {\it between} thermodynamic ensembles can be achieved.
Thus the histogram reweighting technique is very general.

Of course, there are limitations to this technique. For instance, one cannot use data obtained at $\beta$ to
obtain accurate values of $\langle O\rangle$ at \emph{all} $\beta'$.
The general rule is that one is limited to reweighting to parameters close to that at which the simulation
was performed: in the above example one would only be able to use histogram reweighting to obtain reliable estimates 
of $\langle O\rangle$ at $\beta'$ close to $\beta$.

The toolkit supports histogram reweighting for a wide range of thermodynamic parameters and ensembles.
To elaborate, the toolkit supports reweighting operations of the form
\begin{equation}
w_i=\exp\Bigl[-a(b'-b)s_i+\eta_i\Bigr],
\end{equation}
where $a$ is some constant, $b'$ is the new value of the thermodynamic parameter to be reweighted, $b$
is the value of the parameter used in the simulation, $s$ is the physical observable coupled to
the ensemble parameter, and $\eta_i$ is an optional bias applied to configuration $i$ -- something used
in FED methods (see Section~\ref{sec:fed}).
The user must specify the observables which $a$, $b$, $b'$ and $s$ correspond to such that the desired 
reweighting operation is performed -- as well as provide $\eta_i$ for all $i$ if applicable. For
example, choosing $a$ to be the constant 1, $b$ and $b'$ to correspond to the inverse temperature, 
$s$ to correspond to the energy of the system, and ignoring $\eta_i$ (or, equivalently, setting $\eta_i=0$
for all $i$), one recovers Eq.~\ref{reweight_w}, i.e. 
the operation corresponding to reweighting the inverse temperature.



\begin{figure}
\begin{center}
  \includegraphics[width=0.7\textwidth,angle=270]{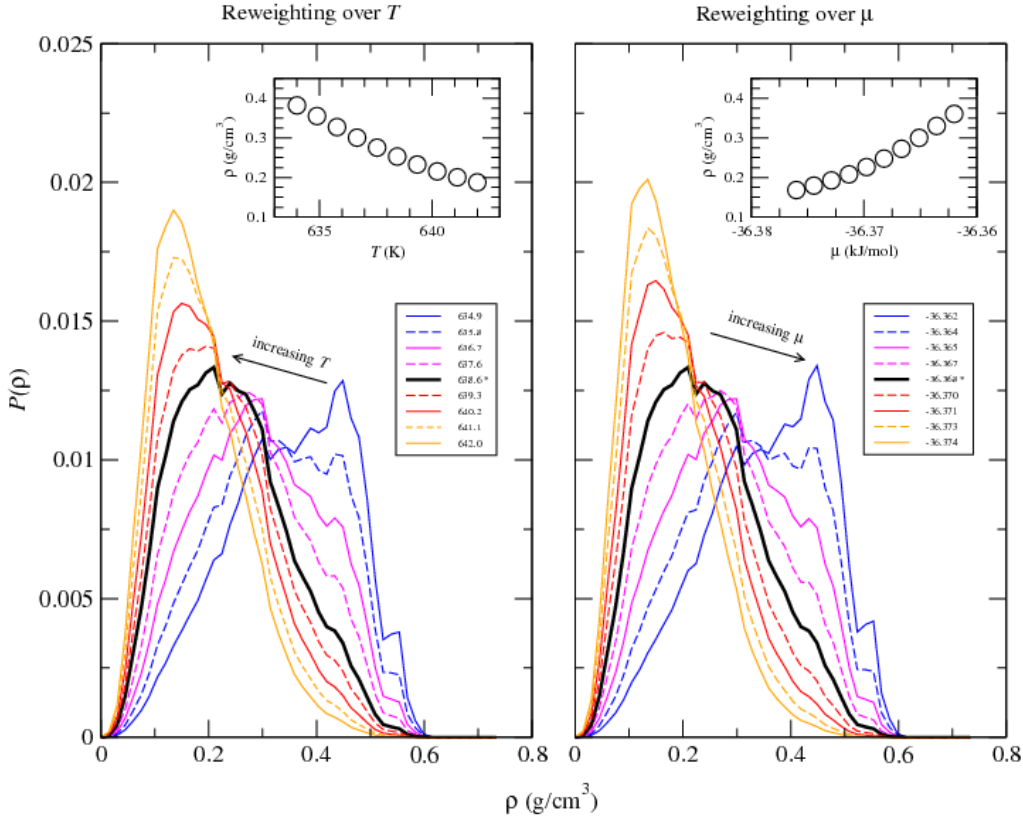}
\caption{Results of applying the histogram reweighting toolkit to data obtained from a \DD 
GCMC simulation of SPC/E water at temperature 638.6~K and chemical potential -36.368~kJ/mol. The left-hand panel shows 
the effect of reweighting the simulation data, specifically, the density probability distribution 
function, to different temperatures. The thick black curve corresponds to the `raw' simulation data, 
while the thin solid and dashed curves are probability distributions at nearby temperatures -- as 
labeled -- obtained by reweighting. The inset shows the corresponding mean density vs. temperature. 
Similar applies to the right-hand panel, but instead instead the data is reweighted to different chemical potentials.}
\label{SPCE_reweighting}
\end{center}
\end{figure}

Fig.~\ref{SPCE_reweighting} provides example results obtained using the toolkit. Here, the toolkit 
has been used to reweight data obtained from a \D $\mu VT$ simulation of SPC/E water~\cite{Berendsen_1987}
near the critical point to nearby temperatures $T$ and chemical potentials $\mu$. The simulation was performed at 
$T=$~638.6 K and $\mu=-36.368$ kJ/mol (where $\mu$ here is the excess chemical potential, without a 
correction added to account for self-polarisation~\cite{Berendsen_1987}), using a simulation box with
dimensions 20\AA~$\times$~20\AA~$\times$ 20\AA. The cut-off for both the Lennard-Jones interactions and
the real-space electrostatic interactions was 6\AA; no tail corrections were applied to the Lennard-Jones 
interactions; the Ewald summation parameter $\eta$ was 0.54722; and a cut-off of 0.55\AA$^{-1}$ used 
for the reciprocal-space component of the electrostatic energy. Moreover, the simulation length was
30,000,000 MC moves, where the relative proportions of different MC move types was 
insert:delete:translate:rotate = 50:50:256:256.
As can be seen in the figure, increasing $\mu$ has the effect of skewing the density probability 
density function towards higher densities, with the opposite effect when decreasing $\mu$ -- as
expected. Similarly, increasing $T$ skews the density probability distribution towards lower 
densities.

We emphasise that the histogram reweighting functionality within the toolkit is not limited to data
obtained from \D simulations. The toolkit could be used to apply histogram reweighting to other sources
of data, e.g. data output by other molecular simulation programs. This would entail writing a 
`plug-in' class for the toolkit. 
Moreover the toolkit can be used to perform \emph{multiple histogram reweighting}~\cite{Ferrenberg_1989}, 
in which data obtained from multiple simulations performed 
at different, say, inverse temperatures $\beta_1,\beta_2,\beta_3,\dotsc$, is reweighted simultaneously 
to calculate observables at an inverse temperature $\beta'$ not probed directly by any of the simulations. 
Multiple histogram reweighting is an extremely powerful method, allowing one to interpolate the value of
observables \emph{between} thermodynamic parameters used in simulations. A closely-related technique is the 
weighted histogram analysis method (WHAM), which we apply later in Section~\ref{sec:applications}.

\subsection{Accessing and using the toolkit}
The Python toolkit is provided alongside \D on CCPForge (see Section~\ref{sec:access}).
The package is agnostic to the choice of Python 2 or 3, and does not depend on any
packages which are not widely available. Instruction on how to use the toolkit is provided in the form of Jupyter
notebooks distributed with the toolkit, as well as documentation embedded in the source code according to
standard Python practices.

\section{Free energy difference (FED) calculations}\label{sec:fed}

Free energy is a function of the thermodynamic state and, as such, it serves as a measure of the thermodynamic stability of a system under given external and internal constraints. That is, the most stable state always has the lowest free energy. Written formally as a function of the parameters that are associated with the constraints, the free energy landscape fully describes thermodynamic equilibria and metastability, corresponding to the global and local free energy minima, respectively. On the other hand, free energy maxima represent thermodynamic barriers and, hence, (reversible) work done on pathways connecting different states of the same system. Therefore, knowing the free energy dependence on one or more parameters $\{q\}$ (often called `order parameters' or `reaction coordinates'), is of great importance and help in studying phase stability, phase transformations, molecular aggregation and many other phenomena in soft and condensed matter.

For example, the liquid-vapour surface tension of a fluid can be obtained from the free energy profile over the density (i.e. with
$q$ as the density). Another example is the free energy associated with a small molecule adsorbing to a surface, or binding to a large
molecule (e.g. a protein). This could be obtained from the free energy profile over the distance of the small molecule from the
binding site. (This is similar to our example in Section~\ref{sec:applications_lipid}).

The free energy profile, $F(q)$, along a given parameter, $q$, can be expressed via the corresponding probability distribution, $P(q)$:
\begin{equation}\label{Fq_def}
\beta F(q) = -\ln P(q) + \text{const.},
\end{equation}
where $\beta\equiv 1/(k_BT)$, $T$ is the temperature, $k_B$ is Boltzmann's constant, and the arbitrary constant reflects
the fact that it is only free energy \emph{differences}, not absolute free energies, which are physically significant.
As mentioned in Section~\ref{sec:intro}, standard MC and MD simulations sample configurations from a thermodynamic ensemble
which describes the `real' system of interest. Such simulations could therefore be used to measure $P(q)$, and 
hence $F(q)$. 

However, in many cases the standard `brute-force' sampling in $q$ space is hindered by either entropic bottlenecks (e.g. in crystal formation/transformation, protein folding) and/or high energy barriers (due to strong energetic coupling, e.g. in self-organised (bio-) molecular aggregates). In such cases, concerned with rare events and long relaxation times, the probability of spontaneous transitioning between minima on the free energy landscape is very low, which makes it practically impossible to obtain $P(q)$ reliably in the relevant $q$-range from an unbiased simulation.

FED methods seek to address this problem. In these methods a \emph{bias} is added to the sampling in order 
to `cancel out' the effect of the free energy barrier. The bias is realised by adding an extra contribution to the Hamiltonian of the 
system, which depends on $q$. We denote this contribution as $U_b(q)$. If $U_b(q)$ is judiciously chosen, then the free energy 
barrier is `canceled out', allowing the simulation to sample the entire range of $q$ space of interest in a reasonable simulation time.
Of course, modifying the Hamiltonian in this way means that $P(q)$, and hence the corresponding 
$F(q)$ (see Eq.~\ref{Fq_def}), obtained from the simulation will not reflect the actual Hamiltonian we are interested in. 
Rather the probability distribution obtained from the \emph{biased simulation} will be the \emph{biased probability distribution} 
$P_b(q)$, as opposed to the \emph{unbiased probability distribution} $P_u(q)$ which we are actually interested in -- and which is
related to the `true' free energy profile via Eq.~\ref{Fq_def}. 
Fortunately $P_u(q)$ can be recovered from $P_b(q)$ because we know the bias $U_b(q)$ used to modify the Hamiltonian. 
The relevant equation is
\begin{equation}\label{pq_pb}
P_u(q)\propto P_b(q)\exp\bigl[\beta U_b(q)\bigr].
\end{equation}
With this in mind the expected value of any observable $O$ can also be obtained from the biased simulation via
\begin{equation}
\langle O\rangle_u=\frac{\sum_iO_ie^{\beta U_b(q_i)}}{\sum_ie^{\beta U_b(q_i)}},
\end{equation}
where $\langle O\rangle_u$ is the expected value of $O$ for the unbiased Hamiltonian; $O_i$ is the observable corresponding 
to configuration $i$, which has order parameter $q=q_i$; and the sum over $i$ is over configurations sampled in the \emph{biased}
simulation.

Thus FED methods entail adding a bias to the Hamiltonian in order to cancel out a free energy barrier, enabling the whole range
of $q$ space to be sampled efficiently, and then removing the effects of the bias in post-processing by exploiting the fact that the bias
is known (Eq.~\ref{pq_pb}).

In practice, however, a crucial problem is that an acceptable $U_b(q)$, i.e. one which sufficiently cancels out the free energy barrier,
is not known from the outset. The different FED methods amount to different approaches to solving this problem, i.e. different
methods for obtaining $U_b(q)$. \D implements the most commonly-used FED methods, which we now describe. For a more thorough discussion of free energies and FED
methods see, e.g.~\cite{Bruce_2004}.

\subsection{Umbrella sampling}
\emph{Umbrella sampling} (US) under a harmonic bias ({\it harmonic US or HUS} hereafter) is one of the oldest FED approaches~\cite{Torrie_Valleau_1974,Torrie_Valleau_1977}.
%
%
Nowadays HUS simulation is a standard protocol typically available in every simulation package, and \DD is no exception. 
Although not a requirement for the method per se, umbrella sampling often employs a harmonic biasing potential,
\begin{equation}
\label{HUS_bias}
U_b(q) = \frac{k_f}{2} (q-q_0)^2,
\end{equation}
where $k_f$ is the force constant and $q_0$ is the parameter value corresponding to the bias minimum. The parabolic form of the bias effectively restricts sampling to a rather narrow $q$-range and, at best, allows one to overcome only one free energy barrier in a single simulation. Therefore, it is a common practice (and the most efficient way of using HUS) to partition the $q$ space into a number of overlapping windows, each being explored by a separate simulation with a window-specific set of $k_f$ and $q_0$. 
From each simulation the free energy profile $F(q)$ for that window can be obtained from the biased probability distribution $P_b(q)$:
\begin{equation}
\label{Fq_bias}
\beta F(q) = -\ln P_b(q) -\beta U_b(q) + \text{const.},
\end{equation}
which follows from Eqs.~\ref{Fq_def} and~\ref{pq_pb}. In practice, $P_b(q)$ is estimated from the histogram of visits over $q$ space, $H_b(q)$, obtained from the simulation: $P_b(q)\propto H_b(q)$.

At the completion of all simulations, the data are to be pooled together to calculate $F(q)$ over the whole range of $q$ space. Crucially, the $F(q)$ for each window is only determined up to an {\it arbitrary} constant. 
These constants are initially determined by the normalization factors for $H_b(q)$ in each window, meaning that the portions of $F(q)$ are shifted with respect to one other by arbitrary amounts. Therefore, it  is necessary to optimally combine the data by ``stitching together'' the FE portions corresponding to separately sampled $q$-windows. 
There exist established methods for `stitching' the $F(q)$ from all windows to obtain the total $F(q)$~\cite{Ferrenberg_1988,Kastner_2011,Shirts_2008}. The most popular is the {\it weighted histogram analysis method} WHAM)~\cite{Ferrenberg_1989,Kumar_1992,Souaille_2001}, and it is standard practice to calculate free energy profiles with the aid of WHAM post-processing utilities. A utility for WHAM post-processing is also provided with the \D package, and its use is demonstrated in Section~\ref{sec:applications_lipid}.

The main advantage of HUS over more modern FED methods (described below) is its stability, owing to the use of an analytical biasing function that does not vary in the course of simulation.
A well-behaved continuous biasing potential is preferable for reliable free energy estimates along `viscous' reaction coordinates for which diffusion in the order parameter can be variable, with fast and sluggish regions~\cite{Brukhno_2008,Tian_2014} (which is detrimental to the convergence of iterative methods for bias optimization). Such an example is discussed in Section~\ref{sec:applications_lipid}.
However, calculating free energy profiles within the US framework over a broad range of $q$ values, and especially in the vicinity of free energy barriers (usually the most interesting regions), invariably requires tedious {\it trial and error} simulations in many overlapping windows in order to determine the optimal values of $k_f$ and $q_0$ for each window, which are unknown in advance. It is for this reason that more modern FED methods, which automate the process of determining the optimal bias $U_b(q)$, are superior for sampling over large regions of $q$ space (where the diffusion properties of $q$ allow for that).

It is worth noting that, apart from the analytical harmonic potential, \D can work with arbitrary biasing forms provided as numerically tabulated input. This feature also facilitates the possible use of US in production runs following optimization of a tabulated bias with the aid of other FED techniques described below.

\subsection{Expanded ensemble method}
Different variants of the \emph{expanded ensemble} (EE) method~\cite{Lyubartsev_1992,Broukhno_2000}, and a number of other similar schemes (like the multicanonical ensemble~\cite{Berg_Neuhaus_1992}, simulated tempering~\cite{Marinari_Parisi_1992}, and the flat histogram method~\cite{WangJS_2000}) were devised with the aim of alleviating the limitations of HUS.
While in the HUS method the bias function is fixed, in the EE method the bias function is iteratively updated in a set of relatively short simulations, thereby `learning' the optimal bias form for overcoming the free energy barriers in $q$ space. This enables free energy profiles to be evaluated in significantly broader ranges of order parameter than would be feasibly possible with HUS. 

The EE method exploits the fact that, if a bias were used 
which would yield a flat histogram $H_b(q)$ over $q$ space (in other words, \emph{uniform sampling} over $q$ space), then Eq.~\ref{Fq_bias} would become
\begin{equation}\label{Fq_flat_hist}
F(q) = -U_b(q) + \text{const.},
\end{equation}
since $H_b(q)$ is a constant for all $q$. Thus, the optimum bias potential should perfectly compensate for the underlying free energy profile.
The problem of determining $F(q)$ is therefore equivalent to the problem of determining a $U_b(q)$ which yields a flat histogram.

To this end the EE method employs a self-consistent iteration for automatically updating the bias $U_b(q)$ starting with some initial guess $U_b^{(0)}(q)$ (usually $U_b^{(0)}(q)=0$ for all $q$):
\begin{equation}\label{EE_bias}
\beta U_b^{(k+1)}(q) = \beta U_b^{(k)}(q) + \lambda \ln{H_b^{(k)}(q)},
\end{equation}
where $k$ is the iteration number and $\lambda \in (0,1]$ is an adjustable {\it feedback factor} allowing control of the convergence of the bias updating procedure. Clearly, at each iteration $k$ a separate simulation is performed with the current bias, $U_b^{(k)}(q)$, and Eq.~\ref{EE_bias} is used for updating the bias before the next ($k+1$) iteration.

In practice, to optimise the numerical stability of the algorithm, the last term in Eq.~\ref{EE_bias} is normally replaced by $\ln{(H_b^{(k)}(q) / H_{ref}^{(k)})}$, with $H_{ref}^{(k)}$ being some reference value which ensures that the biasing function is kept within reasonable bounds. Noting that, $q$ space is discretised into $M_q$ bins for the purpose of evaluating the histogram $H_b^{(k)}(q)$ and tabulating the bias function $U_b(q)$, sensible choices for $H_{ref}^{(k)}$ include: the maximum or minimum number of visits for any bin in $H_b^{(k)}(q)$; or the expected number of visits to any bin in the case of a flat histogram, i.e. $N_s^{(k)}/M_q$, where $N_s^{(k)}$ is the number of samples considered in iteration $k$ (or, equivalently, the total number of visits over all bins in $H_b^{(k)}(q)$). In \D we opted for a slightly different approach. Except for the case when $q$ is the center-of-mass separation (see below), upon updating the bias using Eq.\ref{EE_bias}, \D merely subtracts its maximum value, $U_{b,max}(q)$, so that the current free energy estimate (i.e. $-U_b(q)$) is always kept positive with its global minimum equal to zero. In the case of center-of-mass separation, however, the natural `zero' for the free energy profile is at infinite separation where all the intermolecular interactions vanish. In simulations this limit is rarely reached, but it is nevertheless natural to level the free energy profile such that its long-range tail, corresponding to large separations, is zero. Accordingly, for the case of center-of-mass (COM) separation, \D subtracts from Eq.\ref{EE_bias} the average value of $U_b(q)$ obtained from the $10$ visited bins with the largest separations (assuming $M_q>>10$).

\subsection{Wang-Landau algorithm}
The \emph{Wang-Landau} (WL) scheme was originally suggested for the calculation of density of states (or entropy)~\cite{Wang_2001}, and later adapted for free energies. In this method the system is constantly pushed away from areas of $q$ space which have already been sampled during the simulation. This is achieved by \emph{continuously} updating the biasing function $U_b(q)$. 
In effect, the tabulated bias gradually takes on the shape of the underlying free energy landscape (with the negative sign), which results in progressive flattening of the histogram over $q$ space $H_b(q)$. 
When $H_b(q)$ becomes sufficiently uniform, $F(q)$ can be approximated by $U_b(q)$ via Eq.~\ref{Fq_flat_hist}. 

The WL bias update procedure is as follows. After every MC attempt on variation of the order parameter $q$, the biasing potential for the {\it resulting} value of $q$ is updated,
\begin{equation}\label{WL_bias}
\beta U_b(q)\leftarrow \beta U_b(q)+\Delta,
\end{equation}
where $\Delta$ is an adjustable parameter $\leq k_BT$. 
Note that $\Delta>0$, and so the above update corresponds to making $q$ \emph{less favourable} in the future with regards to the sampling. Clearly, with large values of $\Delta$ the WL method is capable of driving the system through $q$ space very efficiently. However, such an `overrun', albeit readily producing a flat histogram, does not guarantee acceptable precision in the free energy profile (rather the opposite). In effect, $\Delta$ determines the lower bound for the precision in the resulting free energy profile and, hence, is to be gradually reduced in a series of iterations. It is common to start with relatively large initial value, say $\Delta=1$, to accumulate very rough estimates of $U_b(q)$ in a rather short simulation, then decrease $\Delta$, run another, preferably longer, simulation and proceed in this manner until a 
satisfactorily refined $U_b(q)$ is obtained. The criteria for `a satisfactorily refined' bias are: (i) a sufficiently small $\Delta$ value and (ii) a flat histogram generated by that value. However, for complex systems obtaining a sufficiently flat histogram can be extremely time consuming due to intricate hysteresis in sampling. It is then advisable, upon reaching an acceptable convergence (i.e. obtaining an acceptably uniform histogram), to fix the bias and perform a long productive simulation and then use Eq.~\ref{Fq_bias}.

\subsection{Practical remarks on the EE and WL iterations.} 
The EE and WL protocols share one common feature - an iteration is required for bias refinement. A typical uninitiated iteration in both cases starts with relatively short simulation runs providing initial rough estimates for the biasing function. As the iteration progresses the length of refining runs has to be increased. Finally, a production simulation with the obtained well-refined bias may be needed. In \DD this staged iterative protocol is implemented internally, so the user can easily run the entire iteration in one go (see the \DD manual for details).

\subsection{Transition matrix}
The aim of the EE and WL methods is to determine the `ideal' bias function $U_b(q)$ which yields a flat histogram $H_b(q)$. 
In the \emph{transition matrix} (TM) method~\cite{Smith_1995,Fitzgerald_1999} the 
aim is the same, but the approach is very different. The ideal bias function is related to the \emph{unbiased} probability distribution via
\begin{equation}\label{TM_Ub}
U_b(q)=-F(q)=\frac{1}{\beta}\ln P_u(q),
\end{equation}
which follows from Eqs.~\ref{Fq_def} and \ref{Fq_flat_hist}. 
In the TM method one logs the observed transitions between regions of $q$ space during the simulation, 
accumulating the information in a \emph{collection matrix} $H_u(q,q')$. This matrix is used to calculate $P_u(q)$ (see below), and then 
$U_b(q)$, via the above equation. $U_b(q)$ is calculated from $H_u(q,q')$ in this manner continuously throughout the simulation: as the simulation
proceeds, and more transitions are logged in $H_u(q,q')$, the estimate of $U_b(q)$ becomes increasingly accurate, until eventually 
$U_b(q)$ yields uniform sampling over $q$ space.

To elaborate, $H_u(q,q')$ is a count of the number of transitions observed to have occurred from $q$ to $q'$  \emph{for the unbiased 
Hamiltonian}. One \emph{could} (though this is not done in practice, see below) accumulate $H_u(q,q')$ by using the following update procedure 
in an \emph{unbiased} simulation: for every attempted MC move which takes the system from $q$ to $q'$, perform the update 
\begin{equation}
H_u(q,q')\leftarrow H_u(q,q')+1
\end{equation}
if the move is accepted and 
\begin{equation}
H_u(q,q)\leftarrow H_u(q,q)+1
\end{equation}
if the move is rejected (in which case the system remains at $q$ after the move, which corresponds to a transition from $q$ to $q$). 
However this is only applicable in an unbiased simulation, and hence is unsuitable for our purposes.
Fortunately there is a generalisation of the above procedure which can be used to obtain $H_u(q,q')$, \emph{even in a biased simulation}: for 
every attempted MC move which takes the system from $q$ to $q'$, perform both the updates
\begin{equation}\label{TM_update_1}
H_u(q,q')\leftarrow H_u(q,q')+p
\end{equation}
and
\begin{equation}\label{TM_update_2}
H_u(q,q)\leftarrow H_u(q,q)+(1-p)
\end{equation}
regardless of whether the move is accepted or rejected, where $p$ is the probability that the move would be accepted \emph{if there were no 
biasing}. Note that $p$ is trivial to calculate in a simulation which uses biasing.

In the TM method $H_u(q,q')$ is updated every MC move using this procedure, continually accumulating information about the nature of the 
transitions over $q$ space for the unbiased Hamiltonian, even though the dynamics of the system is governed by a biased Hamiltonian.
With $H_u(q,q')$ obtained in this manner, one then estimates the \emph{transition matrix} $T_u(q,q')$, where $T_u(q,q')$ is 
the conditional probability of the system transitioning to order parameter $q'$ given that it currently has order parameter $q$. The relevant 
equation is
\begin{equation}
T_u(q,q') = \frac{H_u(q,q')}{\sum_{q''}H_u(q,q'')}.
\end{equation}
$T_u(q,q')$ is then itself used to calculate $P_u(q)$ by solving the detailed balance equation
\begin{equation}\label{tm_detailed_balance}
T_u(q,q') P_u(q) = T_u(q',q)P_u(q').
\end{equation}
(See the user manual for technical details on how this is done in \Dns). Finally, as mentioned above, $U_b(q)$ is obtained from 
$P_u(q)$ using Eq.~\ref{TM_Ub}. 

Note that $U_b(q)$ is updated often enough during the simulation that, in effect, $U_b(q)$ always reflects the `up-to-date' matrix $H(q,q')$, 
and hence the `best possible guess' for the ideal $U_b(q)$ given the information gathered so far during the simulation.

The TM method has proved extremely efficient, especially for systems exhibiting steep free energy barriers. 
One reason for its efficiency is that no information is ever `thrown away'. All information regarding the \emph{unbiased} movement
across $q$ space that could be obtained from the simulation so far is `stored' in $H_u(q,q')$, and all of this information
is folded in to the bias.
Another pleasing aspect of the TM method is that it parallelises well. One can partition $q$ space into windows, assign different
simulations to accumulate collection matrices $H_u(q,q')$ for each of these windows, and then pool these matrices together, using the
resulting `total' collection matrix to evaluate the total bias function $U_b(q)$ over all $q$ space. \D supports this methodology.
In fact this is how $U_b(q)$ was evaluated in the lattice-switch MC study presented later in Section~\ref{sec:applications_lsmc}.

\subsection{Testing the overall integrity of MC calculations}

As software developers, we have to pay great attention to the correctness and accuracy of the core algorithms implemented in \Dns~2. For example, it is crucial to ensure self-consistency of the energy (re-)calculation routines which are many and specific for every type of MC move. To this end, \DD has a very powerful, yet conceptually simple, ``first-aid'' debugging instrument which automates the flagging of {\it accumulated} errors due to inaccuracies in the updates of various energy contributions. It is called {\it the rolling energy check}. That is, periodically the total energy and all its separate terms are recalculated from scratch and the accumulated (rolling) energies are then checked against these newly recalculated value(s). This procedure relies, of course, on the (presumed) validity of the total energy calculation, which has to be assured only occasionally (say, during major code refactoring or upon introduction of new interaction terms). In practice, this check is carried out at least once -- at the end of a simulation, but one can require more frequent energy checks by specifying in the input the number of MC steps between two consecutive checks.

The ability to evaluate FED profiles versus the particle (or COM) separation enables another powerful technique for testing and ensuring the overall integrity of internal computation workflows in one go, including energy calculations, free energy estimates, and the replica-exchange procedure. 

Consider a system where only two particles are present in the simulation box. In this case a FED calculation with respect to the particle separation has to reproduce the underlying interaction potential or the sum thereof,
\begin{equation}
\beta W(r) = -\ln(p(r)/p(\infty)) = -\ln\langle\exp[-\beta\sum u(r)]\rangle
\end{equation}
where the average reduces to a single value at a given distance $r$. 

\begin{figure}
\begin{center}
\includegraphics[width=0.45\textwidth,angle=0]{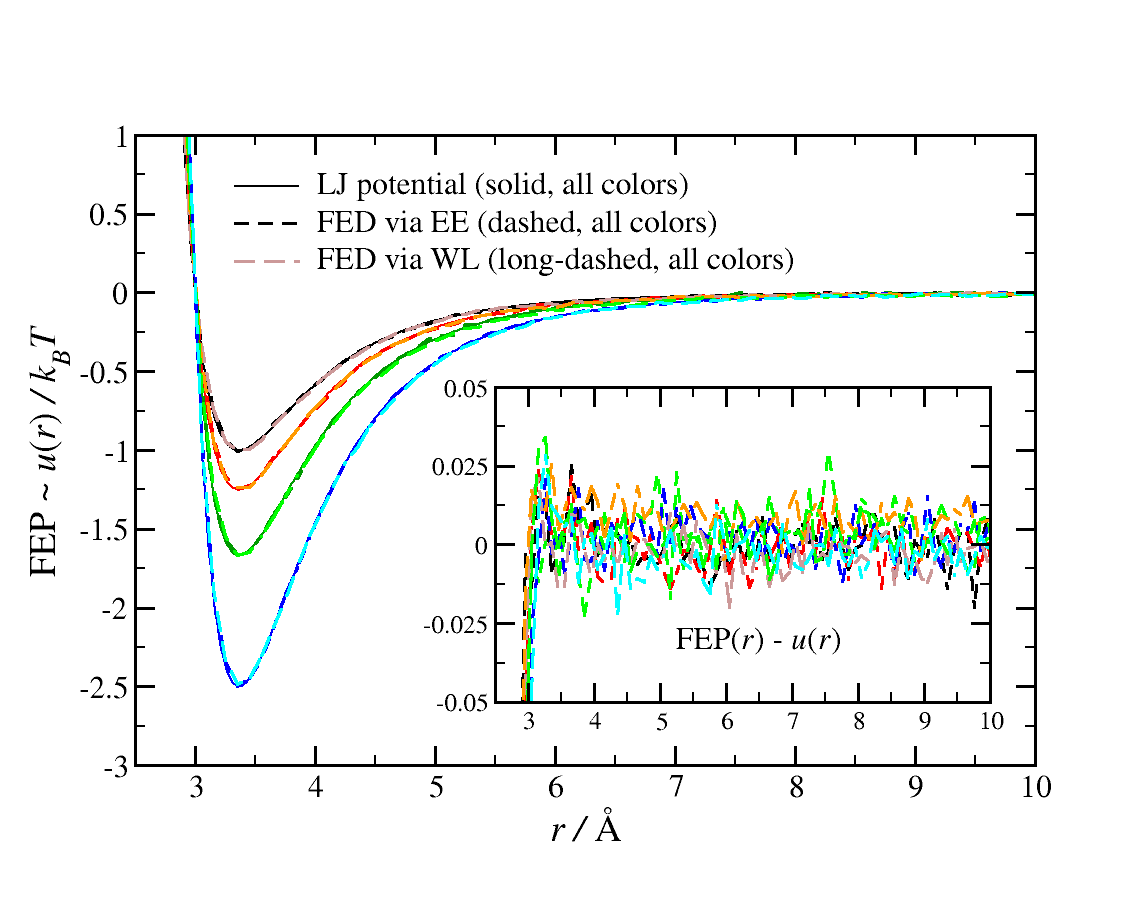}  
\hskip 0.3cm
\includegraphics[width=0.45\textwidth,angle=0]{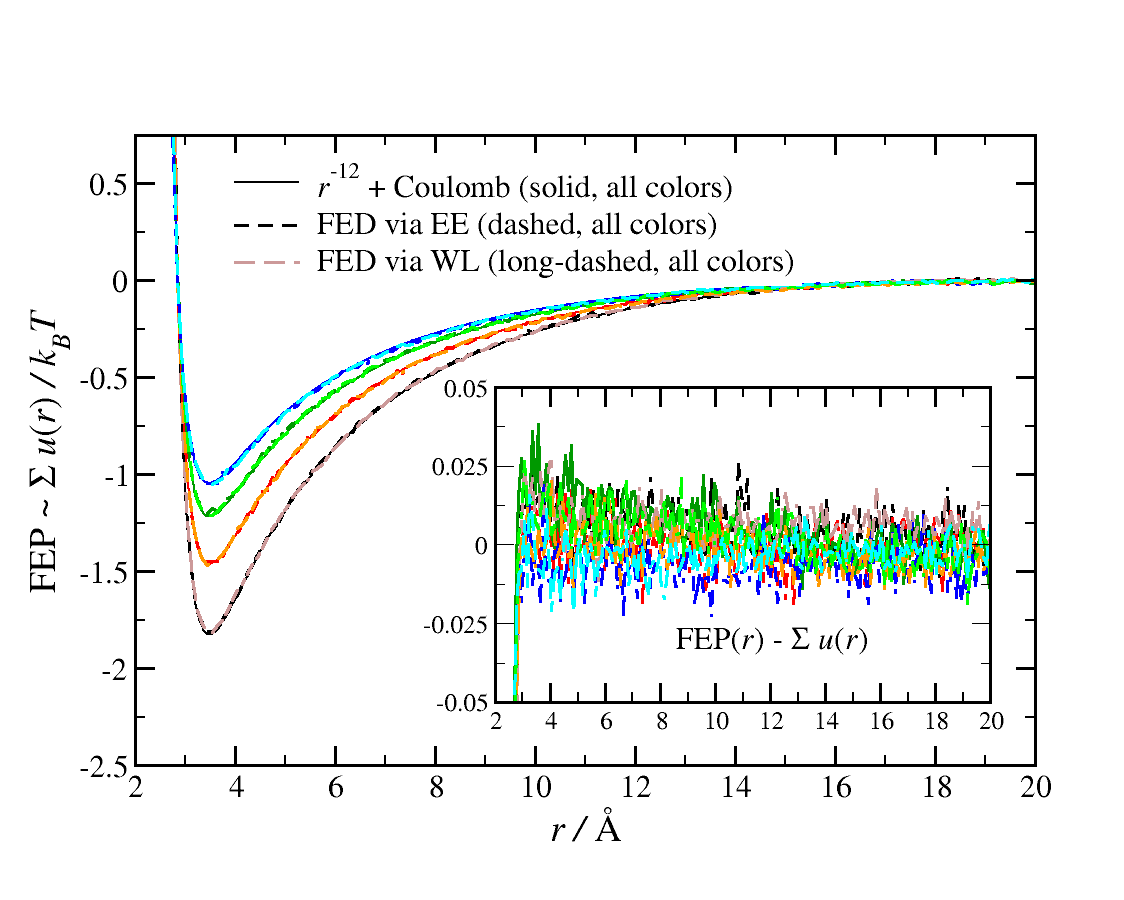}
\caption{Representative test calculations of FE profiles (FEP) reproducing pairwise interactions in two cases: (1) truncated and shifted LJ potential ($\sigma=3$~\AA, $r_{cut}=12$~\AA), (2) sum of repulsive $r^{-12}$ potential and force-shifted (attractive) Coulomb interaction for a pair of oppositely charged ions, $\sigma=3$~\AA, $r_{cut}=20$~\AA. The inserts show the FEP deviations from the analytical functional forms.  In both cases the total number of MC steps (samples) was $16$~million, including preliminary stages of rough estimation (half of the simulation time). Note that the data for a set of 4 temperature values were obtained in a single simulation by employing multicanonical replica-exchange (aka parallel tempering) scheme. }
\label{RE-FED_tests}
\end{center}
\end{figure}

To exemplify, in Fig.~\ref{RE-FED_tests} the obtained free energy profiles (FEP's) vs. particle separation are shown for a pair of interacting particles in the two cases: (i) the pure Lennard-Jones interaction (left-hand panel) and (ii) a combination of repulsive soft core ($A r^{-12}$) and force-shifted Coulomb interactions (right-hand panel; the force-shifted electrostatic potential was chosen for illustrative purposes only, as it smoothly vanishes at the cutoff distance, making it easier to check the FEP against it). In both cases the FED evaluation was combined with periodic replica-exchange configuration swaps within a set of four temperatures. Clearly, the underlying pair interactions are reproduced with high precision (the statistical error in the FEP's is below 0.05 $k_B T$, see the inserts in Fig.~\ref{RE-FED_tests}).

\begin{figure}
\begin{center}
  \includegraphics[width=0.45\textwidth,angle=0]{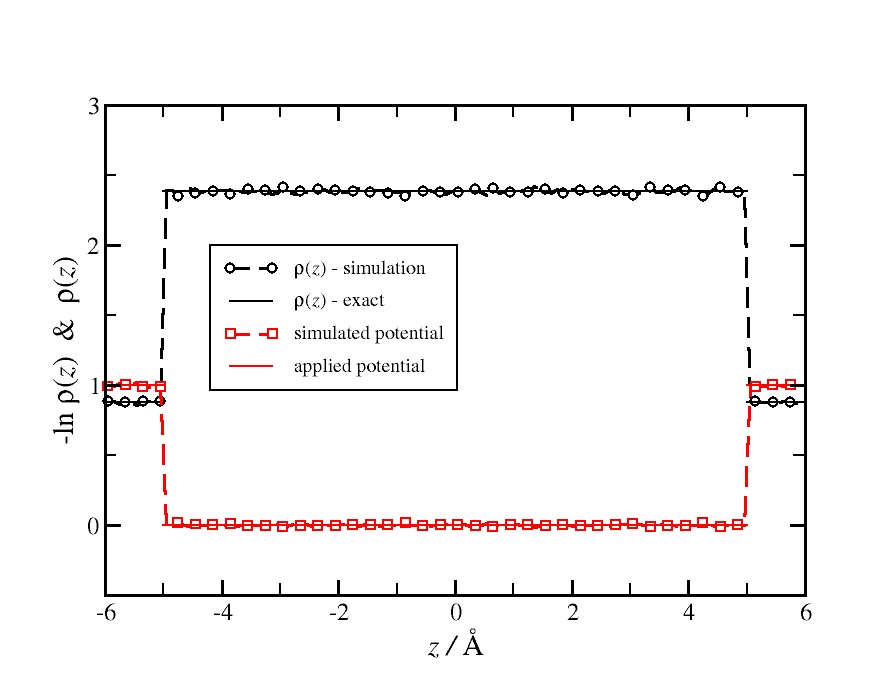}
\hskip 0.4cm
\includegraphics[width=0.47\textwidth,angle=0]{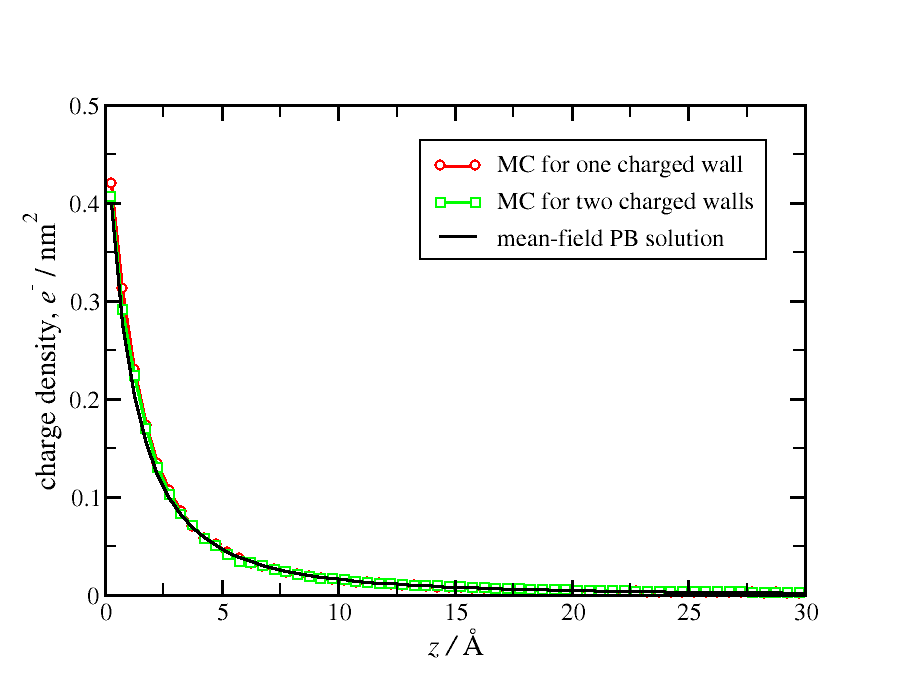}
\caption{Test calculations in the planar pore (slit) geometry. Left panel: density distribution for ideal gas (256 non-interacting particles) constrained to a slit, where the primary cell dimensions are: $11\times 11\times 12$~\AA, and an external potential in the form of a stepwise repulsive shoulder of width $d_w=1$~\AA\ is applied by each of the two flat walls placed at $z_{w}=\pm 6$~\AA. Apart from the density data from MC simulation that are compared with the known exact density levels (black lines), the external potential is shown to be reproduced by $-\ln(\rho(z))$ (red lines). Right panel: simulated density distributions of the surface counterions in a slit with one or two walls bearing a surface charge density of $0.01 e^{-}/$\AA$^2$ ($\Delta z_{w}=200$~\AA) are compared with the corresponding Poisson-Boltzmann (mean-field) solution. Note that dielectric permittivity of water $\epsilon_{w}=78.7$ and the MFA long-range correction for electrostatics outside the primary cell were used in these simulations.}
\label{Slit_tests}
\end{center}
\end{figure}

Two additional examples of test calculations are given in Fig.~\ref{Slit_tests} for the quasi-2D slit geometry. 
This sort of physically meaningful and relatively short simulations (along with their very quick counterparts, so-called regression tests) constitute the core of \DD testing and debugging suite -- an ultimate means for verifying and maintaining the validity of \DD code through its development cycle.

\subsection{Lattice-switch Monte Carlo}

Lattice-switch Monte Carlo (LSMC)\cite{Bruce_1997,Bruce_2000} is a method for evaluating the free
energy difference $\Delta \mathcal{F}=\mathcal{F}_1-\mathcal{F}_2$ 
\footnote{Here we use the symbols $F$ to denote a Helmholtz free energy, $G$ to denote a Gibbs free
energy, and $\mathcal{F}$ to denote a generic free energy -- either Helmholtz or Gibbs.}
between two metastable solid phases 1 and 2.
\footnote{A generalisation of LSMC, named \emph{phase-switch} Monte Carlo, can be used to evaluate the free
energy difference between a solid and a fluid. Note that in the manual, source code and training 
materials for \DD the term phase-switch Monte Carlo is used to refer to all such `switching' Monte
Carlo techniques.}
LSMC has been used to add insight into the phase behaviour of a wide range of systems
(see~\cite{Underwood_2017} for a brief review of previous LSMC applications). Moreover it has
been used to develop force fields which accurately capture the locations of phase transitions
\cite{Mendelev_2016,Quigley_2014}. However, while there have been many applications of LSMC to 
\emph{atomic} systems (including `atomic' soft-matter such as hard spheres), applications to 
\emph{molecular} systems have been few
\footnote{We do not count monatomic water, which was studied with LSMC in~\cite{Quigley_2014},
as a `molecular system' here since monatomic water is an `atomic' force field}.
The only studies we are aware of are~\cite{Marechal_2008} (where LSMC was applied to colloidal hard
dumbbells),~\cite{Raiteri_2010} (calcium carbonate) and~\cite{Bridgwater_2014} (butane). 

To facilitate the uptake of LSMC by the community, especially with regards to molecular systems
modelled by complex force fields, we have implemented LSMC within \Dns. 
Below we provide a brief description of LSMC, followed by a demonstration that \D reproduces 
results in the literature for various fundamental systems. Later, in Section 
\ref{sec:applications}, we apply LSMC to the problem of phase stability of plastic crystal phases in 
the water model TIP4P/2005~\cite{Abascal_2005,Aragones_2009}, in order to demonstrate the applicability
of the method to `realistic' force fields.

\subsubsection{Methodology}
The key feature of LSMC is a `switch' MC move which, used alongside conventional 
Monte Carlo moves (e.g. atom translation, molecular rotation, volume), enables the system to
explore the two solid phases under consideration in a single simulation of reasonable length. This
in turn allows $\Delta \mathcal{F}$ to be evaluated via
\begin{equation}\label{LSMC_DeltaF}
\Delta \mathcal{F}=-k_BT\ln(p_1/p_2),
\end{equation}
where $p_1$ and $p_2$ are the probabilities of the system being in phases 1 and 2 deduced
from the simulation. This approach is not possible with conventional simulation methods on 
account of the large free energy barrier separating the phases, which prevents transitions
between the phases occurring in accessible simulation lengths.

The switch move exploits the fact that, in a solid phase, the particle positions closely 
resemble the ideal crystal lattice which characterises the phase. To elaborate, in a solid the
position of particle $i$ can be expressed as
\begin{equation}
\mathbf{r}_i=\mathbf{R}_i+\mathbf{u}_i,
\end{equation}
where $\mathbf{R}_i$ is the lattice site for particle $i$ and $\mathbf{u}_i$ is the
\emph{displacement} of $i$ from its lattice site. (Note that the displacements are small since 
the positions closely resemble the ideal crystal lattice). In a switch move from phase 1 to
phase 2, the underlying phase-1 lattice is `switched' for a phase-2 lattice, \emph{while 
preserving the particles' displacements}. More precisely, the particle positions are 
transformed from $\mathbf{r}^{(1)}_i\to\mathbf{r}^{(2)}_i$ for all $i$, where
\begin{align}
\mathbf{r}^{(1)}_i&=\mathbf{R}^{(1)}_i+\mathbf{u}_i, \\
\mathbf{r}^{(2)}_i&=\mathbf{R}^{(2)}_i+\mathbf{u}_i,
\end{align}
and $\mathbf{R}^{(1)}_i$ and $\mathbf{R}^{(2)}_i$ are the lattice sites for $i$ in phases 1 and
2 respectively. Note that the transformation yields a `plausible' phase-2 configuration, i.e.
the phase-2 configuration closely resembles the phase-2 ideal lattice. This is the source of 
the success of LSMC: the switch move always attempts to take the system from a configuration in 
the current phase to a plausible configuration in the `other' phase, bypassing the free
energy barrier separating the phases. 

In `atomic' systems, where the particles have no internal degrees of freedom (e.g., 
orientation, bond angles and lengths), the description of the switch transformation given
above is sufficient. However, for molecular systems there is the question of how to
transform the particles' internal degrees of freedom during the switch~\cite{Bridgwater_2014}. 
In this work we only consider molecular systems comprised of rigid molecules, for which the 
\emph{orientations} of the molecules constitute the internal degrees of freedom. For such systems,
\DD employs the following transformation for molecular orientations.
Let $Q^{(1)}_i$ denote the orientation of molecule $i$ in a \emph{reference configuration}
characteristic of phase 1, and similarly for $Q^{(2)}_i$. We shall refer to $Q^{(1)}_i$ and
$Q^{(2)}_i$ as the phase-1 and phase-2 \emph{reference orientations} of $i$. Moreover let 
$\mathcal{R}_i$ denote the rotation required to transform $Q^{(1)}_i$ to $Q^{(2)}_i$:
\begin{equation}
Q^{(2)}_i=\mathcal{R}_iQ^{(1)}_i.
\end{equation}
In \DD the switch move from phase 1 to phase 2 transforms the orientation $q^{(1)}_i$ of 
molecule $i$ as follows: $q^{(1)}_i\to q^{(2)}_i$ for all $i$, with
\begin{equation}
q^{(2)}_i=\mathcal{R}_iq^{(1)}_i.
\end{equation}
Thus in \DD the rotation linking the phase-1 and phase-2 orientations of $i$ is
always the same, and is the rotation $\mathcal{R}_i$ linking the phase-1 and phase-2
reference orientations for $i$.
This mapping of $q^{(1)}_i\to q^{(2)}_i$ is suitable for plastic crystal phases, 
where, by definition, all orientations of a molecule have a reasonably likely probability of
being realised at equilibrium.

In LSMC switch moves are frequently attempted. If they are also frequently successful, the result 
is that the system transitions between the two phases under consideration often, allowing Eq. 
\ref{LSMC_DeltaF} to be used to calculate $\Delta \mathcal{F}$. However, it turns out that, even with 
frequent switch moves, transitions between the two phases are too rarely accepted for Eq. 
\ref{LSMC_DeltaF} to be applied. In effect a free energy barrier separating the phases remains -- 
though it is many orders of magnitude smaller than would be the case without switch moves. 
Fortunately the barrier is small enough that it can be surmounted using FED methods such as those
described earlier in this section. In this case $q$ is the \emph{LSMC order parameter} $M$, 
defined as follows for a configuration $\sigma$:
\begin{equation}\label{LSMC_order_param}
M(\sigma)=
\begin{cases}
-\Delta E(\sigma)\text{ if $\sigma$ belongs to phase 1} \\
+\Delta E(\sigma)\text{ if $\sigma$ belongs to phase 2},
\end{cases}
\end{equation}
where $\Delta E(\sigma)$ is the energy change upon performing a switch move from configuration 
$\sigma$. It turns out that this order parameter does the job of distinguishing both phases,
as well as, when used with switch moves, defining an efficient path between the phases -- see
\cite{Underwood_2017} for more details.

To summarise this section, there are two aspects to LSMC: switch moves, and the use of FED
methods sampling over the aforementioned order parameter. Both aspects lead to the two 
phases under consideration being explored in a single simulation of reasonable length, ultimately
enabling the free energy difference between the phases to be calculated via Eq. 
\ref{LSMC_DeltaF}.

\subsubsection{Validation: results for fundamental systems}

After implementing LSMC in \Dns, our first task was to validate it against predictions of other 
codes and results in the literature for fundamental systems. We used \DD to calculate the 
following:
\begin{enumerate}
\item $\Delta \mathcal{F}$ between the hcp and fcc phases of the hard sphere solid 
($\Delta \mathcal{F}\equiv \mathcal{F}_{\text{hcp}}-F_{\text{hcp}}$) at density $\rho=0.7778
\rho_{\text{cp}}$, 
where $\rho_{\text{cp}}$ is the density corresponding to close packing. This $\Delta \mathcal{F}$ was 
calculated in the $NVT$ ensemble using a system size of $N=216$ spheres. The benchmark 
$\Delta \mathcal{F}$, to which the value obtained from \DD was compared, was taken from~\cite{Bruce_2000}. 
\item $\Delta \mathcal{F}$ between the hcp and fcc phases of the hard sphere solid at pressure 
$P=14.58k_BT/D^3$, where $D$ is the hard-sphere diameter. This $\Delta \mathcal{F}$ was calculated in 
the $NPT$ ensemble using a system size of $N=216$ spheres and `isotropic' volume moves which 
preserve the shape of the system. The benchmark $\Delta \mathcal{F}$ was calculated using the LSMC code 
in~\cite{Underwood_2017}. 
\item $\Delta \mathcal{F}$ between the hcp and fcc phases of the Lennard-Jones solid at pressure $P=0$ 
and temperature $T=0.1\varepsilon/k_B$, where $\varepsilon$ is the well depth of the
Lennard-Jones potential and $k_B$ is the Boltzmann constant. This $\Delta \mathcal{F}$ was calculated in 
the $NPT$ ensemble using a system size of $N=216$ particles and isotropic volume moves. The 
benchmark $\Delta \mathcal{F}$ was was calculated using the LSMC code in~\cite{Underwood_2017}.
\item $\Delta \mathcal{F}$ between the hcp and fcc plastic crystal phases of hard dumbbells. The dumbbell
particles consisted of two intersecting hard spheres of radius $D$, whose centres are separated
by $0.15D$. This $\Delta \mathcal{F}$ was calculated in the $NVT$ ensemble using a system size 
of $N=864$ dumbbells at a density of $\rho^*=1.15$, where $\rho^*\equiv d^3\rho$, where $\rho$ is
the number of dumbbells per unit volume and $d$ is the diameter of a sphere with the same volume 
as the dumbbell. The benchmark $\Delta \mathcal{F}$ was taken from~\cite{Marechal_2008}.
\end{enumerate}
The $\Delta \mathcal{F}$ obtained using \DD are compared against the benchmarks in Table 
\ref{table_LSMC_fund}. In all cases agreement is found between \DD and the benchmarks, 
providing confidence that our implementation of LSMC is correct.

Note that these precise form part of \Dns's test suite, and provide an excellent test of \Dns's functionality beyond just LSMC.

\begin{table}
\tbl{Free energy differences obtained with lattice-switch Monte Carlo in \Dns, and benchmark values, 
for the fundamental systems described in the main text. Quoted errors here reflect standard errors in the mean
obtained from block averaging.}
{\begin{tabular}{cccc}
 \toprule
 Calculation                          & Units                & Benchmark       & \DD \\
 \colrule
  (1) Hard sphere solid, $NVT$   & $10^{-5}k_BT$        & 133(4)          & 137(4)   \\
  (2) Hard sphere solid, $NPT$   & $10^{-5}k_BT$        & 123(6)          & 135(6) \\
  (3) Lennard-Jones solid, $NPT$ & $10^{-3}\varepsilon$ & 1.283(7)        & 1.290(11)  \\
  (4) Hard-dumbbell solid, $NVT$ & $10^{-3}k_BT$        & -5(1)           & -4.3(5) \\ 
 \botrule
\end{tabular}}
\label{table_LSMC_fund}
\end{table}

\section{Scientific applications}\label{sec:applications}
We now present two examples which showcase the FED functionality described in the previous section. 
Specifically, in Section~\ref{sec:applications_lipid} we use umbrella sampling in \D to elucidate the structure of a lipid 
bilayer. Then in Section~\ref{sec:applications_lsmc} we use LSMC to study the stability of
two plastic crystal phases in the water model TIP4P/2005.

Input files for the simulations performed in this section are available on the CCPForge webpage for \D (see Section 
\ref{sec:access}), to serve as a full account of the simulation methodology and to aid users who wish to perform similar simulations.

\subsection{Free energy of a lipid within a bilayer}\label{sec:applications_lipid}

In our first case study we demonstrate the ability of \DD to treat complex molecular systems by employing the recently introduced `\textit{Dry Martini}' force field for DOPC (dioleoylphosphatidylcholine) lipids~\cite{Martini_2015} in simulation of a lipid bilayer - a system typical in biomolecular simulation. The results from \DD simulations are compared to those obtained with the use of two MD packages: DL\_POLY~\cite{dlpoly2} and Gromacs~\cite{Gromacs4_2008}.

The Martini force field represents a set of coarse-grain (CG) models for organic (bio-) molecules, such as hydrocarbons, surfactants, lipids, polysaccharides etc., including polarisable and non-polarisable CG water models. The Dry Martini model, in particular, goes one step further in simplification and removes the aqueous environment from consideration by replacing it with a continuous medium, which greatly reduces the computational demand in simulation of biomolecules. That is, Dry Martini belongs to the type of `implicit solvent' (or `solvent-free') models which are particularly suited for Monte Carlo simulation.

A schematic representation of a DOPC CG lipid and a bilayer is shown in Fig.~\ref{DOPC_bilayer}~(left panel). As with any coarse-grain model, the Martini model lumps together a few atomic groups to form a CG particle (otherwise known as bead or superatom). In particular, we use the following notation for the DOPC lipid CG beads: NC3$^+$ for the positively charged choline group; PO4$^-$ for the negatively charged phosphate group; GLY for the two glycerole beads; CHS and CHD for the tail beads uniting hydrocarbon groups, where `S' and `D' letters distinquish between alkane and alkene types (the former for groups with only single carbon-carbon bonds and the latter for those with one double bond.) Comprehensive details of the force-field can be found in~\cite{Martini_2015} and the references therein.

The initial setup for simulation of a DOPC bilayer was created with the use of the CHARMM-GUI membrane-builder online tool~\cite{CHARMM-GUI}, which can automatically generate the necessary input files in a number of popular formats. We opted to start with the inputs in Gromacs format and then convert the configuration and force-field files to both \DD and DL\_POLY-4 formats to allow, where possible, comparison of the results between the three simulation engines. 

The bilayer structure was assembled from 256 DOPC lipids (128 per leaflet). As is typical in bilayer simulations, the two leaflets and hydrophilic surfaces are, on average, parallel to the XY plane and percolate in the X and Y directions via periodic boundary conditions, see Fig.~\ref{DOPC_bilayer}. The bilayer structure was initially centered at the origin of the simulation cell, which had dimensions $92\times92\times100$~\AA, and briefly equilibrated in the $NVT$ ensemble at 310 K ($t_{eq}=100$ ps, and $N_{eq}=10^5$ MC translation steps per CG particle). Following~\cite{Martini_2015}, with such a setup the natural \textit{tensionless} conditions can be simulated in an isothermal-isotension ($NP_{xy}T$) ensemble where the $Z$ dimension of the simulation box is kept constant and the lateral external pressure $P_{xy}=0$. 

Two types of simulations were carried out: (1) $NPT$ simulations under the aforementioned isothermal-isotension conditions (using all the three packages; note that isotropic $NPT$ was used in the case of DL\_POLY-4), and (2) biased HUS simulations in the $NP_{xy}T$ ensemble aimed at calculating the work done upon reversible translocation of a lipid molecule across and out of the bilayer, i.e. the free energy profile (FEP) for a lipid molecule being driven along the Z axis. In the first instance our aim was to examine the \DD capability to simulate a molecular (bilayer) structure, and compare the observed bilayer properties against those obtained with two renowned MD packages. Then, to further investigate usability of \DD for FED evaluation in complex molecular systems, the FEP obtained for Dry Martini model by using \DD is compared with the data calculated with Gromacs for a widely-used united atom lipid model (known as the Berger force-field) in aqueous environment modelled explicitly with SPC/E water model~\cite{Berger_opls_lipids, DOPC_lipid_sims}.

The main settings for the MD simulations using Gromacs were taken from \cite{Martini_2015} and then closely resembled in DL\_POLY simulations, but also informed the MC setup for \Dns. In particular, all simulations employed the truncated and shifted variants of Van der Waals (Lennard-Jones) and Coulomb interaction potentials, which smoothly vanish at the cutoff, $R_{cut}=12$~\AA. The thermostat and barostat coupling constants $\tau_T = 1.0$ and $\tau_P = 4.0$ ps and the compressibility of $3\times 10^{-4}$ bar$^{-1}$ were used. The main difference between the two MD engines was that DL\_POLY appeared stricter in application of the stochastic (Langevin) thermostat which resulted in 5 times smaller time step required, cf. 4 fs vs 20 fs in DL\_POLY and Gromacs, respectively. Full equilibration with respect to the area per lipid was reached after $t_{eq}^{tot}=2$~ns (MD) and $N_{eq}^{tot}=2$ million sweeps (MC). 

All simulations exploited 16 parallel MPI processes per job and were executed on the SCARF HPC cluster at Rutherford Appleton Laboratory (RAL, STFC)~\cite{SCARF}. In the case of \D we found it optimal for the current bilayer system to run simulations in the 4$\times$4 mode, with 4 independent workgroups involving 4 parallel worker-processes in each (for loop parallelisation), whereby each run generated 4 independent trajectories at the same simulation conditions (see the appendix for more details).

\subsubsection{Bilayer properties and lipid order}\label{Bilayer}

\begin{figure}
\begin{center}
  \includegraphics[width=0.45\textwidth,angle=0]{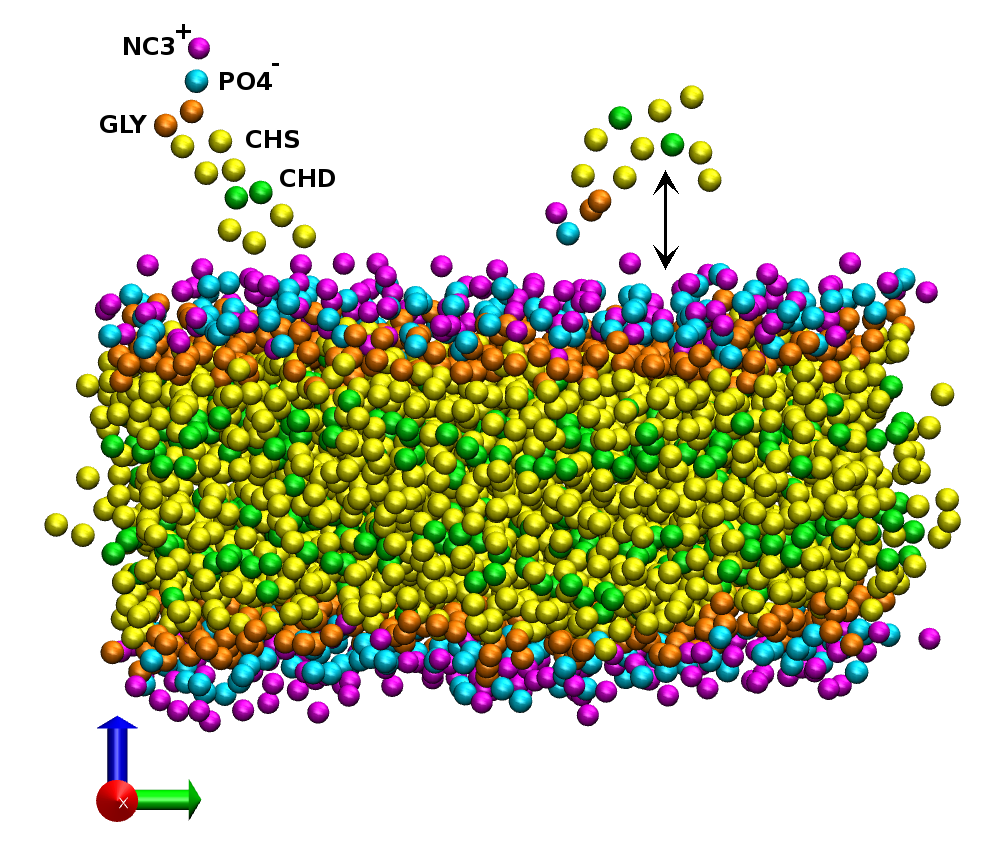}
  \includegraphics[width=0.45\textwidth,angle=0]{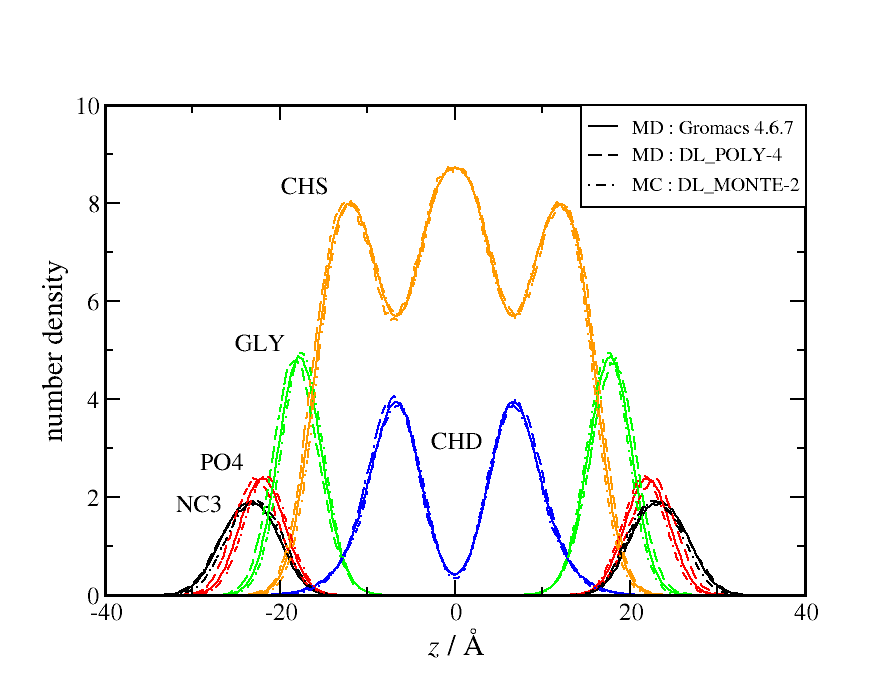}
\caption{Left panel: Visualisation graphic of the bilayer assembled of DOPC CG lipids and simulated by using Dry Martini force-field. The two molecules overlaid above the membrane illustrate two typical lipid configurations in biased simulations where a lipid molecule is pulled in and out the membrane. Right panel: $Z$-density profiles across the bilayer for the monomeric CG units (beads) comprising DOPC lipid molecules. The \DD MC data (dot-dashed lines) are compared with the MD simulation results obtained with Gromacs (solid lines) and DL\_POLY-4 (dashed lines).}
\label{DOPC_bilayer}
\end{center}
\end{figure}

In Fig.~\ref{DOPC_bilayer} (right-hand panel) we compare the density profiles ($z$-density) for different CG beads across the bilayer obtained with the use of the \Dns, Gromacs and DL\_POLY packages. Clearly, the profiles corresponding to the same bead type but obtained with different simulation engines practically coincide, with only marginal variations. Other bilayer properties are compared in Table~\ref{table_bilayer_props}. Viewed altogether, the data allow us to conclude that all the three simulation packages are in a good agreement with each other (as expected). 

It is worth noting that generally bilayers simulated with Martini models are noticeably (about 10\%) thicker and tighter (i.e. more compact in the lateral dimensions) than observed experimentally, cf. the data for POPC lipids in \cite{Martini_2015}. As can be seen from the Table, this is reflected in our results too. On the other hand, the Martini estimated area per DOPC lipid is very close to that obtained with the popular Berger (united atom) model. The 
observed deviations from the experimental values are, of course, the result of a compromise between the model detail and the simulation efficiency. As is reported by Siu et~al,~\cite{DOPC_lipid_sims} fully atomistic models, such as CHARMM-27 and GAFF, perform considerably better in all respects.

\begin{table}
\tbl{Properties of a DOPC lipid bilayer modelled by the Dry Martini force field. The data obtained by DL\_MONTE, Gromacs and DL\_POLY 4 are compared for: bilayer thickness estimated by the distances between the density peaks for NC3 and PO4 beads, $XY$-projected area per lipid, and various $z$-order parameters for lipids within the bilayer, see the text for details. In all figures the standard error (calculated by block-averaging) is contained in the last digit shown. For reference, the membrane thickness and area per lipid are also given for the Berger force-field and from experiment.~\cite{DOPC_lipid_sims} }
{\begin{tabular}{lllllll}
 \toprule
 Property                        & Units         & Gromacs   & DL\_POLY-4   & \DD     & Berger FF & exp.\\
 \colrule
  Bilayer thickness as $d_{NC3}$ & \AA           & 45.9~     & 45.7~        & 45.2~   &               \\
  Bilayer thickness as $d_{PO4}$ & \AA           & 44.0~     & 44.4~        & 43.8~   & 37.2 & 37.1~ \\
  Area per lipid ($XY$-projected) & \AA$^2$       & 65.4~     & 65.9~        & 65.7~   & 66.0 & 72.1~ \\
 \colrule
  Lipid head $z$-angle           & degree        & --        & --           & 45.1$^\circ$~   & 88$^\circ$   & -- \\ 
  Lipid tail $z$-angle           & degree        & --        & --           & 40.6$^\circ$~   & \\ 
  Lipid head $z$-order           & --            & 0.356~    & 0.329~       & 0.327~  & \\
  Lipid tail $z$-order           & --            & 0.417~    & 0.406~       & 0.415~  &  \\
  Bonded CH triplet $z$-order (total)  & --            & 0.335~    & 0.317~       & 0.332~  &  \\
 \colrule
  CHS$_2$ on tail 1 ($sn$1)        & --            & 0.503~    & 0.471~       & 0.488~  &  \\
  CHD$_3$ on tail 1 ($sn$1)        & --            & 0.373~    & 0.367~       & 0.374~  & \\
  CHS$_4$ on tail 1 ($sn$1)        & --            & 0.155~    & 0.147~       & 0.164~  & \\
 \colrule
  CHS$_2$ on tail 2 ($sn$2)        & --            & 0.470~    & 0.452~       & 0.464~  & \\
  CHD$_3$ on tail 2 ($sn$2)        & --            & 0.359~    & 0.338~       & 0.358~  & \\
  CHS$_4$ on tail 2 ($sn$2)        & --            & 0.147~    & 0.129~       & 0.143~  & \\
 \botrule
\end{tabular}}
\label{table_bilayer_props}
\end{table}

Of particular interest is the ordering of lipids within a membrane, as it is characteristic of a specific lipid and determines the phase behaviour of membranes with different composition. Therefore, we performed a comprehensive analysis of the lipid order parameters that are commonly used to characterise their tendency to align. We used the segmental $z$-order parameter as our primary measure,
\begin{equation}
S_z = \frac{1}{2}(3\langle\cos\Theta_{z}\rangle^2 - 1),
\end{equation}
where $\Theta_{z}$ is the angle between the normal to the bilayer surface (approximated by the $Z$-axis) and the vector along a given segment within a lipid molecule. $S_z$ takes on values in the interval [0,1] and directly measures the $z$-alignment of lipid backbone segments. Hence, it is a natural and distinctive parameter for CG lipid models, as opposed to the deuterium order parameter based on carbon-hydrogen alignment which is often used for atomistic models due to having a direct counterpart in experiment, e.g. see~\cite{DOPC_lipid_sims}. Note that $S_z$ can be linked to the deuterium parameter and is estimated to be normally twice the latter~\cite{Akinshina_2016}. The overall $z$-order for bonded carbon-based triplets (see below) is also reported in~\cite{Martini_2015} for POPC CG lipids modelled with the Dry Martini force-field, $S_{z,{\rm POPC}}=0.35$. The several $S_z$ values in Table~\ref{table_bilayer_props} are presented for: lipid head segments NC3$-$GLY~(sn1) (assigned to PO4 beads), full lipid tail 
segments CHS$_1-$CHS$_5$ (assigned to CHD$_3$), and bonded CH triplet segments within tails corresponding to CH$_{k-1}-$CH$_{k+1}$ vectors (assigned to CH$_k$ beads) for each tail. 

First, we see that the average angle of head-group orientation, 45$^\circ$ away from the $Z$-axis in the Martini model, is almost twice as small as the angle reported for the Berger united atom model that predicts a virtually flat orientation of lipid head-groups, 88$^\circ$, i.e. practically parallel to the $XY$-plane. The angle predicted by Martini model appears, though, in better agreement with the most probable head-group orientation angles reported for the all-atom models~\cite{DOPC_lipid_sims}: 59$^\circ$ for GAFF(SPC/E) and 62$^\circ$ for CHARMM-27(TIP3P). Next, we note that the overall $z$-order values for full lipid tails (CHS$_1-$CHS$_5$) are generally higher than those averaged over all bonded triplets (which is also reflected in the corresponding average angles; not shown). Moreover, the $z$-order parameter for bonded triplets varies depending on the location of a given triplet on each lipid tail (see also~\cite{DOPC_lipid_sims, Akinshina_2016}) and drops from approximately 0.48 (CHS$_2$), through 0.36 (CHD$_3$), down to 0.14 (CHS$_4$), where the more abrupt second drop can be attributed to a kink angle of 120$^{\circ}$ between the CHD$_3$ and CHS$_4$ beads (mimicking the effect of a C=C bond). There is also an obvious systematic trend of $S_z$ values being slightly higher for the $sn$1 tail, i.e. the tail that is directly linked to the lipid head-group through a \textit{single} GLY bead (via PO4$-$GLY~($sn$1) bond). 

To summarise, our $S_z$ data indicate that the \textit{overall} $z$-alignment of lipid tails in a membrane is more accurately characterised by the $z$-order of full (CHS$_1-$CHS$_5$) tail vectors, as opposed to the total average $S_z$ over bonded triplets, the only used in~\cite{Martini_2015}. On the other hand, a comprehensive analysis of $S_z$ values, for every bonded triplet (and possibly every CG bond) on each lipid tail, allows for acquiring a detailed picture of the variations in $z$-alignment both between and within lipid tails.

\subsubsection{Evaluating the free energy profile for a lipid pulled across the bilayer} 

Calculation of the potential of mean force (PMF) acting on the center of mass (COM) of a molecule traversing through a biological membrane is a traditional means to study net interactions within membranes, as well as membrane permeability to intra- and extra-cellular agents.~\cite{Jambeck_2015,Huang_2015,Lyu_2017} As an illustrative example of such a calculation, we use \D to evaluate the PMF, or F($z$), for a lipid molecule reversibly translocated across one of the bilayer leaflets. Apart from illustrating the applicability of the program to this end, this case study also aims to test the Dry Martini CG model against the more detailed Berger united atom force-field combined with the SPC/E water model. 

We employ harmonic umbrella sampling (known as `harmonic restraint' in molecular dynamics) in both MC and MD simulations, where the \D MC engine is used for the Dry Martini DOPC model and Gromacs is exploited for atomistic MD simulations. The biasing potential, Eq.~\ref{HUS_bias}, is applied along the $Z$ axis, i.e. it acts selectively on the $z$-component of the COM separation between the restrained lipid molecule and the bilayer, which we denote from here on by $Z_{\rm lip}$ (defined relative to the bilayer mid plane). Considering the very restricted lipid motion across the bilayer and, hence, extremely long relaxation times for a lipid driven out of its natural equilibrium position within the bilayer, several simulations in a set of subranges of $Z_{\rm lip}$ (windows) are necessary in order to equilibrate the system under the influence of the bias in each window and collect sufficient statistics for reliable determination of F($Z_{\rm lip}$). To this end, we use equidistant placement of the bias minima, $Z_0^{(k)}$ ($k$ being the window index), with a step of $2$~\AA~in both the MC and MD simulations. The bias force constant was set equal in all windows, $k_f=4$~$k_B T$~\AA$^{-2}$, which is sufficiently high to restrain the biased lipid diffusion within a window, yet low enough to allow for acceptable overlaps in the probability distributions $P_k(Z_{\rm lip})$ between the neighbouring windows. 

\begin{figure}
\begin{center}
  \includegraphics[width=0.75\textwidth,angle=0]{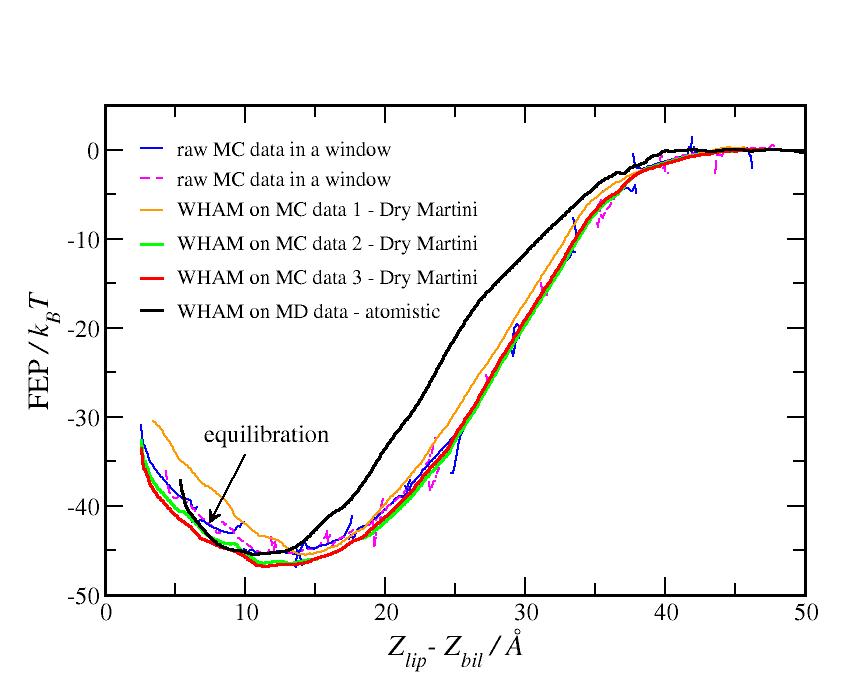}
\caption{Potential of mean force, or FEP($\Delta Z$), for a (DOPC) lipid pulled across a bilayer by means of harmonic umbrella sampling in several (22) overlapping windows. The slow equilibration process and the importance of using the WHAM procedure for optimally combining raw FED data over all windows are emphasised: compare the intermediate raw data (solid blue and dashed magenta lines) with the curves produced by WHAM. Two sets of simulations comprising $16$ million MC sweeps (evenly distributed between 4 parallel sub-processes) were carried out in each of the 22 umbrella windows during equilibration stage (orange and green lines), and the production simulation was twice as long (red line). The black line represents the reference FEP data obtained in MD simulations (Gromacs) for the atomistic DOPC (Berger) model; only the production results from the last 40 ns in each window are shown.}
\label{DOPC-FED-WHAM}
\end{center}
\end{figure}

Four relaxed configurations generated previously in the unbiased (production) $NP_{XY}T$ simulation (Section~\ref{Bilayer}) served for seeding as starting configurations for biased simulations. The windows were populated by performing two preparatory simulations in which the bias minima were set to $Z_0=2$ and $40$~\AA, respectively, whereby providing a strong pull away from the initial $z$-position of the driven lipid molecule. Then, configurations from within the vicinity of each $Z_0^{(k)}$ were extracted from the preparatory trajectories and used as seeds in different umbrella windows. Fig.~\ref{DOPC-FED-WHAM} presents all the obtained F($Z_{\rm lip}$) data, from where it is evident that three subsequent MC simulations were necessary in each window to, first, equilibrate the system (two equilibration runs, 16 million MC sweeps each) and then accumulate sufficient statistics in the production runs (32 million sweeps). A similar equilibration procedure was also required in MD simulations for the atomistic model, which amounted to 20 ns equilibration and 40 ns production runs in all windows (22 in total in both MC and MD cases).

The overall FE profiles were obtained with the aid of a stand-alone WHAM utility (written in Python and provided with \Dns; the `gmx wham' tool was used in the case of Gromacs). That is, the raw (biased) piecewise probability distributions were self-consistently reweighted and combined into the total (de-biased) distribution, which was then converted into the free energy data. For comparison, we also include the raw FED MC results calculated by Eq.~\ref{HUS_bias} in each window after the second equilibration stage. The corresponding FEP fragments were `stitched' together in a plotting software by shifting them with respect to each other along the abscissa axis until an acceptable matching was achieved. We see that, in contrast to this tedious procedure, the WHAM method not only automatically finds the optimum shifts for seamless stitching of the FEP portions, but also effectively smooths out all the spikes and roughness in the overlapping regions between the windows (owing to undersampling at the edges of each window). 

Regarding the comparison of the solvent-free Dry Martini CG model and the significantly more detailed atomistic model, Fig.~\ref{DOPC-FED-WHAM} leads us to two main conclusions. (1) As expected, the overall shape of the free energy profiles obtained with the two models is very similar. The evident discrepancies are mostly observed in the location and width of the global minima in F($Z_{\rm lip}$). In the atomistic model the equilibrium position of a lipid is closer to the bilayer center, and the lipid motion in the $Z$ direction is more hindered as compared to the coarse-grain model ($Z_{\rm eq}\approx 11\pm 5$~\AA~vs $12.5\pm 6.5$~\AA, respectively, within a threshold of $4$~$k_BT$ above the FEP minimum). This is in accord with the aforementioned tendency of increased bilayer thickness observed with Martini force-field. (2) The FEP depth and its slope associated with pulling the lipid out of the 
bilayer are reproduced well by the Dry Martini model. In particular, the discrepancy between the two models in the estimated partitioning free energy, i.e. the difference in the depth of F($Z_{\rm lip}$), lays within 5$\%$, which should be regarded as remarkably good agreement, taking into account the dramatic departure in detail between the two representations.

\subsection{Thermodynamic stability of plastic crystal phases in TIP4P/2005 water}\label{sec:applications_lsmc}


After successfully testing LSMC in \DD for fundamental models (see Section~\ref{sec:fed}), 
the next step was to apply LSMC to molecular systems modelled by realistic force fields. 
Hence we chose to examine the stability of the bcc vs. fcc plastic crystal phases of TIP4P/2005 water 
\cite{Abascal_2005} at $T$=440~K and $P$=80~kbar, with the aim of comparing our results to 
those of Aragones and Vega (AV)~\cite{Aragones_2009} using the 
thermodynamic integration method~\cite{Kirkwood_1935}. This is our second example application of \Dns.

TIP4P/2005~\cite{Abascal_2005} is a rigid model for water in which each molecule is comprised of
4 sites: an O atom, which interacts with O atoms in other molecules via a Lennard-Jones 
potential; two H atoms, each with charge +0.5564$e$ (where $e$ is the proton charge); and an 
additional site named `M', located close to the O atom, which houses the remaining charge in the 
molecule -1.1128$e$. (See~\cite{Abascal_2005} for further details regarding TIP4P/2005).
Note that TIP4P/2005 is of comparable complexity to other `realistic' force fields typically
used in simulations involving small molecules. Hence our forthcoming results serve to illustrate
that LSMC could be used to examine phase stability in molecular crystals modelled with
realistic force fields. One particularly interesting prospect is to use the method to examine the 
phase stability of crystals of small pharmaceutical molecules, such as paracetamol.

We calculated the energies and densities of the bcc and fcc phases (denoted $E_{\text{bcc}}$,
$E_{\text{fcc}}$, $\rho_{\text{bcc}}$ and $\rho_{\text{fcc}}$), as well as the Gibbs free energy 
difference between the phases $\Delta G\equiv (G_{\text{bcc}}-G_{\text{fcc}})$, using LSMC in the $NPT$ 
ensemble.
The mapping from particle positions in the bcc phase to the fcc phase and vice versa used in the LSMC switch
move was the same as used in~\cite{Underwood_2015}.
We considered various system sizes: $N=250$, 432 and 686, where $N$ denotes
the number of molecules in the system. In all of these calculations isotropic volume moves which preserve
the shape of the system were used. 
However, we additionally performed calculations at $N=250$ and 432 in which volume moves were disabled within each 
phase, such that all configurations explored within the bcc and fcc phases corresponded to fixed
densities $\rho_{\text{bcc}}$ and $\rho_{\text{fcc}}$ -- which were chosen before the simulation. 
In these calculations, the system could only change its density/volume upon switching from one phase to another, 
i.e. jumping from density $\rho_{\text{bcc}}$ to $\rho_{\text{fcc}}$ upon a successful switch move from a bcc 
configuration to a fcc configuration or vice versa.
Note that the probability of such a change in density/volume is dependent on the specified pressure in the 
usual manner (see, e.g.~\cite{Frenkel_Smit}). 
We refer to the thermodynamic ensemble sampled in these calculations as the $NPT\rho_1\rho_2$
ensemble, where $\rho_1$ and $\rho_2$ are the chosen densities for phases 1 and 2. 
Crucially, the $NPT\rho_1\rho_2$ becomes equivalent to the $NPT$ ensemble in the limit of large $N$, so long 
as $\rho_1$ and $\rho_2$ correspond to the true equilibrium densities of both phases at the specified $T$ and $P$. 
Hence, for large enough $N$, one will retrieve the correct  $\Delta G$ from a $NPT\rho_1\rho_2$ LSMC calculation
employing the correct densities for both phases.
The reason we consider the $NPT\rho_1\rho_2$ ensemble is that it a closer analogue to the calculation
of $\Delta G$ performed by AV -- to which we will compare our results -- than an LSMC calculation of $\Delta G$ in
the `unconstrained' $NPT$ ensemble (i.e. the $NPT$ ensemble in which the densities of both phase can vary during the
simulation).
AV calculated $\Delta G$ as follows. First, they calculated the equilibrium densities $\rho_{\text{bcc}}$ and
$\rho_{\text{fcc}}$ for each phase using conventional MC simulations. Then, systems corresponding to each phase 
were set up at these densities, and thermodynamic integration was applied \emph{in the $NVT$ ensemble} to 
calculate the \emph{Helmholtz} free energies for each phase, $F_{\text{bcc}}$ and $F_{\text{fcc}}$. Finally, AV
applied the equation $G=F+PV=F+PN/\rho$, using the aforementioned $F_{\text{bcc}}$, $F_{\text{fcc}}$, $\rho_{\text{bcc}}$ 
and $\rho_{\text{fcc}}$, to obtain the \emph{Gibbs} free energies $G_{\text{bcc}}$ and $G_{\text{fcc}}$ for both phases,
from which $\Delta G$ follows trivially. Thus AV's calculation of $\Delta G$ in fact corresponds to the
$NPT\rho_1\rho_2$ ensemble.
Our $NPT\rho_1\rho_2$ calculations utilised the same $\rho_{\text{bcc}}$ and $\rho_{\text{fcc}}$ as AV to allow a
like-for-like comparison as much as possible. In a similar vein, we used the same cut-offs for the Lennard-Jones potential
in the TIP4P/2005 model and real-space part of the Ewald summation as AV.

\begin{figure}
\begin{center}
  \includegraphics[width=0.7\textwidth]{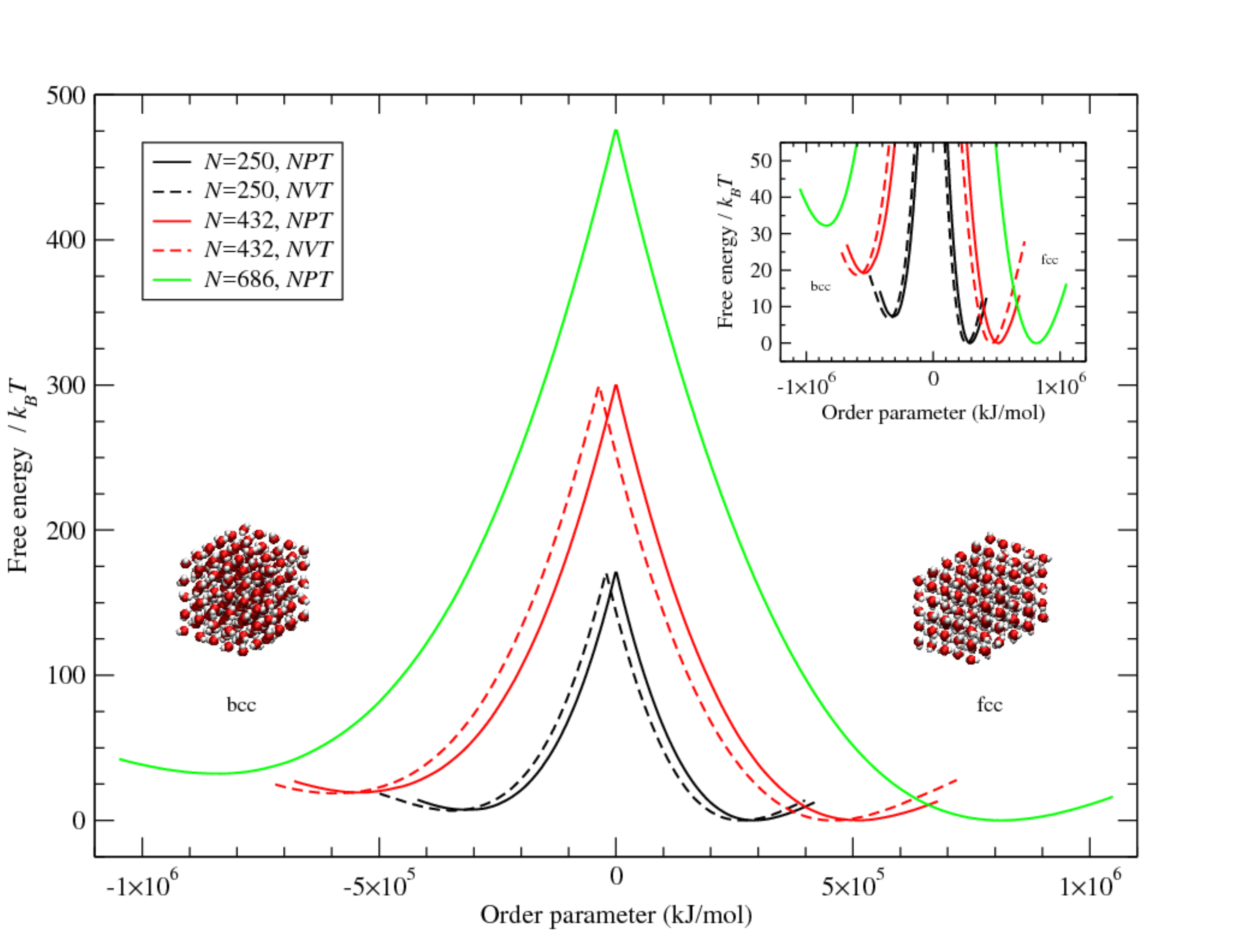}
\caption{Free energy profiles vs. order parameter for LSMC investigations of bcc and fcc plastic crystal phases of TIP4P/2005 water 
at $T$=440~K and $P$=80~kbar. The local minima at negative (positive) order parameters correspond
to the equilibrium states in the bcc (fcc) phase -- as indicated by the labels. Also shown are the reference configurations 
for the $N=250$ simulations for each phase. Note that the fcc configuration is a body-centred tetragonal representation
of fcc~\cite{Underwood_2015}.}
\label{LSMC_weight_functions}
\end{center}
\end{figure}

We calculated the bias functions $U_b(M)$ (where recall that $M$ is the LSMC order parameter, Eq.~\ref{LSMC_order_param})
to use in our LSMC production simulations using the transition matrix (TM) method 
(see Section~\ref{sec:fed}). To elaborate, we first, for each considered system, performed conventional (i.e. non-LSMC)
unbiased MC simulations for each phase, tracking the range of $M$ exhibited by each phase, in order to
deduce an appropriate range of $M$ to consider in our LSMC simulations. We then partitioned $M$ space into 6 windows, and performed 6 
TM simulations, each with the system confined to a different window, in parallel. 
At the completion of
these simulations, the transition matrices obtained for all windows were combined into a global transition matrix, which was 
then used to obtain an estimate of the `ideal' $U_b(M)$ corresponding to uniform sampling over all $M$ space. 

Recall that a bias function $U_b(M)$ which results in uniform sampling is related to the underlying
free energy profile $F(M)$ via $F(M)=-U_b(M)$ (up to an additive constant) (Eq.~\ref{Fq_flat_hist}). The $F(M)$ implied by
the $U_b(M)$ we obtained for each of our considered systems, as described above, are presented in Fig. 
\ref{LSMC_weight_functions}. Note that for all systems the basin in the free energy profile corresponding to the fcc equilibrium 
configurations is lower than the bcc basin. This suggests that the fcc phase is more stable than bcc at the considered $T$ and $P$,
regardless of system size and ensemble.

\begin{figure}
\begin{center}
  \includegraphics[width=0.7\textwidth,angle=270]{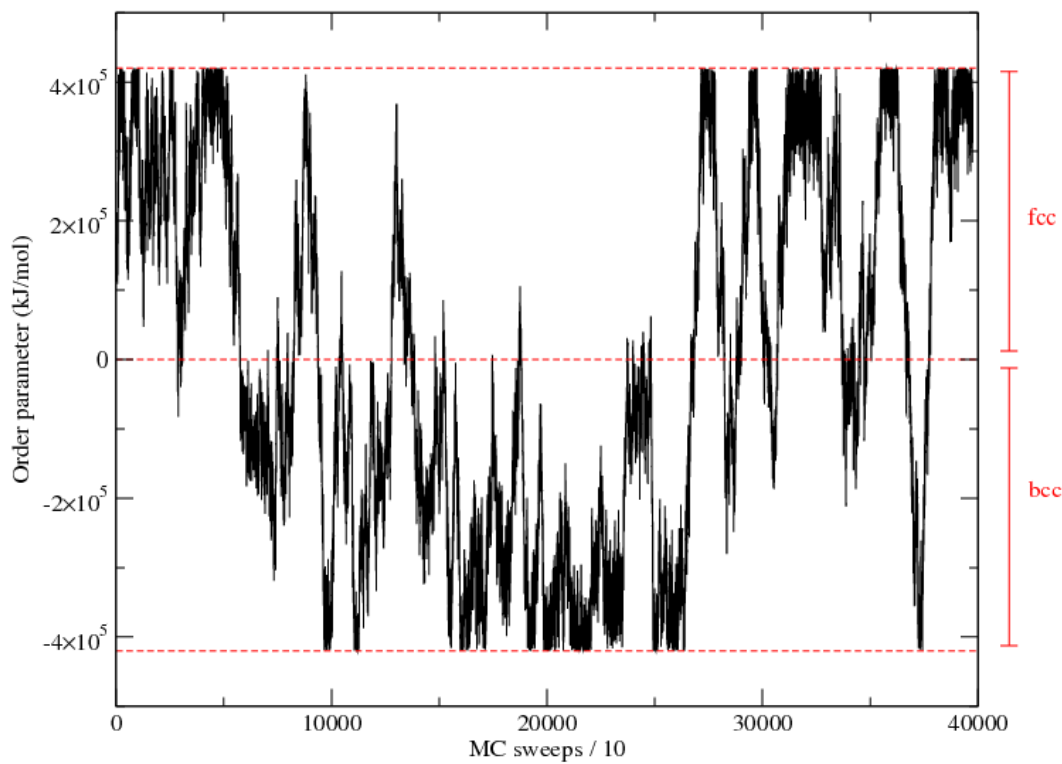}
\caption{Representative trajectory of system through order-parameter space for LSMC simulations of bcc and
fcc plastic crystal phases of water modelled by TIP4P/2005 at $T$=440K and $P$=80kbar. This trajectory 
corresponds to $N=250$ in the $NPT$ ensemble, for which a Monte Carlo 'sweep' is defined as 750 Monte Carlo 
moves, where the relative frequencies of the different types of moves is translation:rotation:switch:volume 
= 250:250:250:1. The regions of order parameter space associated with the bcc and fcc phases are indicated, 
as are the upper and lower bounds of the considered order parameter range (the uppermost and lowermost 
dashed lines). }
\label{LSMC_trajectory}
\end{center}
\end{figure}

Confirmation that this is indeed the case can be found in the results of
our production LSMC simulations. These are presented in Table~\ref{table_LSMC_water}. Moreover a representative
trajectory in order parameter space from a production simulation is given in Fig.~\ref{LSMC_trajectory}; note that both
phases are explored in a single simulation. As can be seen from the table, 
$\Delta G\equiv (G_{\text{bcc}}-G_{\text{fcc}})>0$ for all our calculations, indicating that fcc is the
preferred phase. 
Note also that our $NPT$ results for $\rho_{\text{bcc}}$, $\rho_{\text{fcc}}$, 
$E_{\text{bcc}}$ and $E_{\text{fcc}}$ are all converged with respect to $N$ by $N=250$: the smallest 
system we considered $N=250$ is sufficient to get correct values for these quantities. 
Similar applies to our $NPT\rho_1\rho_2$ results: there is no significant change in the energies
and densities upon moving from $N=250$ to $N=432$. On the other hand, for both $NPT$ and $NPT\rho_1\rho_2$ 
there is a significant change in $\Delta G$ between $N=250$ and $N=432$: the finite size effects are
stronger in the free energy than the single-phase quantities. Unfortunately, for the $NPT$ ensemble, while 
$\Delta G$ at $N=423$ and $N=686$ are in agreement, our $\Delta G$ at $N=686$ lacks the precision to 
conclusively determine whether or not $\Delta G$ has converged by $N=686$ to a precision of more than
0.01$k_BT/N$.

How do our results compare with those of AV? The densities obtained by AV
were $\rho_{\text{bcc}}=1.662$~gcm$^{-3}$ and $\rho_{\text{fcc}}=1.679$~gcm$^3$, which recall are the densities 
we employed in our $NPT\rho_1\rho_2$ calculations. As can be seen from Table~\ref{table_LSMC_water}, our $NPT$
calculations yielded a bcc density which is in excellent agreement with AV. However, the fcc density we obtained
is slightly lower than AV's value, by $\approx 0.004$~gcm$^{-3}$. It is not clear whether or not this discrepancy
is significant, since AV did not report uncertainties for their densities. Similar applies to their value of 
$G_{\text{bcc}}$ and $G_{\text{fcc}}$, and hence $\Delta G$. To double-check that the discrepancy in $\rho_{\text{fcc}}$
was not caused either by a bug in \Dns, or an artifact arising from the choice of unit cell
(our LSMC calculations did not utilise a `conventional' unit-cell representation of fcc, but instead utilised a body-centred
tetragonal representation -- see~\cite{Underwood_2015}), we performed additional 
\emph{conventional} $NPT$ MC calculations (i.e. not LSMC calculations) of various fcc systems with various system sizes 
and cell shapes using \Dns. However the $\rho_{\text{fcc}}$ obtained from these calculations was the same as the $NPT$ LSMC
calculations (not shown).

Regarding the free energy difference, AV found that $\Delta G/(Nk_BT)=0.56$. By contrast
we obtained 0.03(1) from our $NPT$ ensemble calculations ($N=686$) and 0.044(5) from our 
$NPT\rho_1\rho_2$ calculations ($N=432$). As discussed earlier, we expected that our values of $\Delta G$ would
agree with that of AV, especially our $NPT\rho_1\rho_2$ value, since this calculation is closer in spirit to AV's 
calculation than our $NPT$ calculations. It is thus concerning that our $\Delta G$ is more than an order of magnitude
smaller than~\cite{Aragones_2009}. In searching for the source of the discrepancy, we noticed that the two values of 
$G_{\text{fcc}}$  quoted by AV are not self-consistent with their quoted values of
$F_{\text{fcc}}$ and $\rho_{\text{fcc}}$, i.e. $G_{\text{fcc}}\neq F_{\text{fcc}}+PN/\rho_{\text{fcc}}$. 
Using their quoted values of $\rho_{\text{fcc}}$ and $F_{\text{fcc}}$, by our calculation AV's values of 
$G_{\text{fcc}}/(Nk_BT)$ should be 18.60 for the fcc system and 18.57 for the `fcc*' system in, as opposed to 
19.23 and 19.21 respectively.
(By contrast we find that AV's quoted value for $G_{\text{bcc}}$ \emph{is} self-consistent with their quoted values
for $\rho_{\text{bcc}}$ and $F_{\text{bcc}}$).
In light of this, however, the discrepancy between our $\Delta G$ and that of AV widens:
the `corrected' AV value is $\Delta G/(Nk_BT)=1.17$, which is even further from our values, i.e. 0.03(1) from our
$NPT$ calculations and 0.044(5) from our $NPT\rho_1\rho_2$ calculations.
The reason for this is not clear, and requires further investigation. One possibility is that the discrepancy is due
to differences in the implementation of the Ewald summation between us and AV, e.g. we may have used a different cut-off
radius in $k$-space for the reciprocal part of the summation. 

\begin{table}
\tbl{Results of \D LSMC simulations involving the bcc and fcc plastic crystal phases of TIP4P/2005 
water at $T$=440 K and $P$=80 kbar for various ensembles and system sizes. The
significance of the $NPT\rho_1\rho_2$ ensemble is described in the main text. 
For the $NPT\rho_1\rho_2$ simulations the densities of the bcc and fcc phases were fixed 
at 1.662 g/cm$^3$ and 1.679 g/cm$^3$ respectively, as signified by the italicised densities in
the table. Uncertainties reflect standard errors in the mean obtained by block averaging.}
{\begin{tabular}{ccccccc}
 \toprule
 Ensemble   & $N$   & $\rho_{\text{bcc}}$ (g/cm$^3$)  & $\rho_{\text{fcc}}$ (g/cm$^3$) & $E_{\text{bcc}}/N$ (kJ/mol) & $E_{\text{fcc}}/N$ (kJ/mol) & $\Delta G /(Nk_BT)$  \\
 \colrule
  $NPT$     & 250   & 1.6625(1)                     & 1.6750(1)                      & -37.42(1)                     & -36.57(1)                   & 0.0289(7) \\
  $NPT$     & 432   & 1.6623(3)                     & 1.6758(2)                      & -37.41(2)                     & -36.568(8)                  & 0.047(1) \\
  $NPT$     & 686	& 1.6618(2)                     & 1.6747(8)                      & -37.42(1)                     & -36.61(3)                   & 0.03(1)    \\
  $NPT\rho_1\rho_2$     & 250	& \emph{1.662}                  & \emph{1.679}                   & -37.421(8)                    & -36.434(7)                  & 0.0279(9) \\    
  $NPT\rho_1\rho_2$     & 432	& \emph{1.662}                  & \emph{1.679}                   & -37.43(2)                     & -36.43(4)                   & 0.044(5)  \\
 \botrule
\end{tabular}}
\label{table_LSMC_water}
\end{table}

\section{Summary}\label{sec:summary}

We have presented \Dns, a software package for performing Monte Carlo (MC) simulations. \D is open source, and can be obtained from 
CCPForge as described in Section~\ref{sec:access}.
The package is general-purpose in that it includes a wide range of force fields, enabling it to simulate a broad range of systems 
with many MC methods. As well as `standard' MC techniques -- namely, the ability to simulate atomic and molecular systems in 
the canonical ($NVT$), isobaric-isothermal ($NPT$) and grand-canonical ($\mu VT$) ensembles -- various advanced methods are also 
implemented in \Dns. These methods include replica exchange, Gibbs ensemble MC, the ability to treat systems confined to a planar 
pore (i.e. `slit' or `slab' boundary conditions), lattice-switch MC for evaluating free energy differences between polymorphs, and 
various \emph{free energy difference} (FED) methods for evaluating free energy profiles (namely, umbrella sampling, expanded 
ensemble, Wang-Landau, and the transition-matrix method). 
Moreover \D comes with a Python toolkit for managing simulation workflows and applying the histogram reweighting analysis method to
output data.

We have provided an overview of these features of \Dns, paying particular attention to the free energy methods, i.e. 
lattice-switch MC and the FED methods. We have provided two `real world' examples to elucidate the use of these methods in \Dns. 
Specifically, we have applied umbrella sampling to calculate the free energy profile associated with the transolcation of a lipid
through a bilayer; and we have employed lattice-switch MC to examine the thermodynamic stability of two competing plastic crystal 
phases of a water model at high pressure.

Future development of \D will involve further optimisation to improve performance of the program, as well as the addition of new 
functionality deemed to be of value to the community. Alongside this, we plan to expand our existing set of tutorials and examples
in order to improve the usability of the package and facilitate its uptake. We believe \D
will prove useful to practitioners of molecular simulation in a broad range of fields, especially in tackling problems where
MC methods (including MC advanced methods) are the most suitable.

\section*{Acknowledgements}
This work made use of the Balena High Performance Computing Service at the University of Bath.
Computing resources were also provided by STFC Scientific Computing Department's SCARF HPC cluster~\cite{SCARF}.
The support of CCP5~\cite{ccp5} is gratefully acknowledged, as are valuable discussions with Graeme Ackland.

\section*{Funding}
This work was supported by the Engineering and Physical Sciences Research Council (EPSRC) under Grant EP/M011291/1.
Kevin Stratford was funded under the embedded CSE programme of the ARCHER UK National Supercomputing Service 
(\url{http://www.archer.ac.uk}): project eCSE04-4.

\appendix

\section{Performance and optimisation of \Dns}\label{sec:performance}
In this appendix we discuss technical features of \D which relate to its performance.

\subsection{Parallelization mechanisms in \D} 

\begin{figure}
\begin{center}
  \includegraphics[width=0.7\textwidth,angle=0]{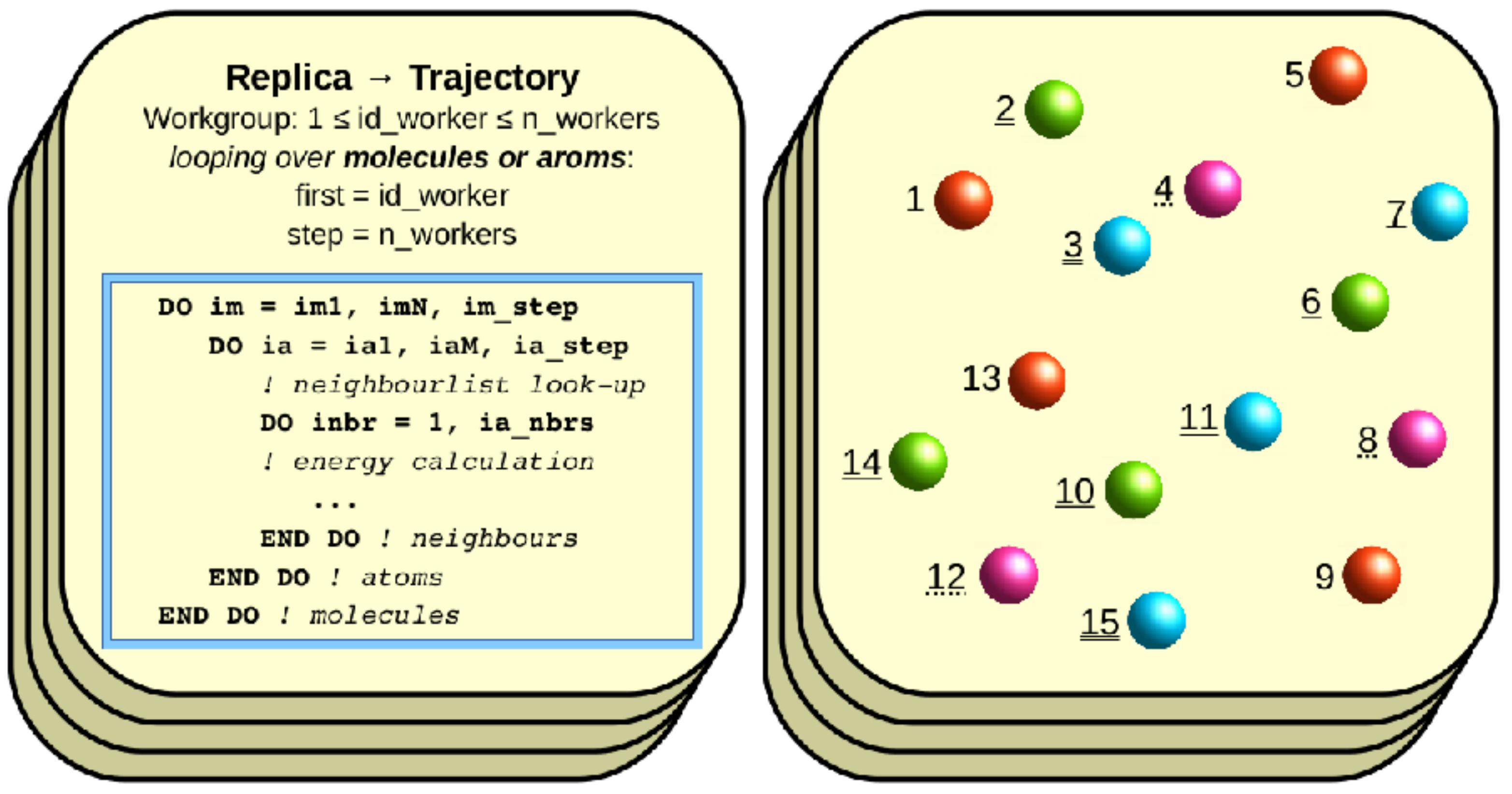}
\caption{Illustration of core loop parallelization within a workgroup, i.e. a group of parallel (MPI) processes working on the same replica 
(configuration), in \Dns. The overlaying layers in the picture signify ``task farming'', with several replicas being treated during the same 
simulation run and replica exchange (pairwise swapping of the simulation box contents) can be periodically attempted in the case of different 
replica temperatures. In the right-hand panel the loop splitting (or \textit{particle decomposition}) is visualised for a system of 15 particles
where particles treated by different ``workers'' (MPI processes) within a workgroup are distinguished by both color and index  underscores 
(4 workers are assumed).}
\label{RE-workgroups}
\end{center}
\end{figure}

DL\_MONTE implements a combination of two approaches to distributing and performing calculations in parallel (e.g. in a HPC environment): 
(1) loop splitting and, thereby, parallelisation of the core routines for energy calculations, which are then carried out jointly by a few MPI 
processes within a so-called ``workgroup''; and (2) internal ``task farming'' by splitting the simulation job between a few ``workgroups'', which 
results in simultaneous generation of several MC trajectories in the course of a single parallel run where the trajectories may either be 
completely independent or periodically exchange configurations (see Fig.~\ref{RE-workgroups}). The latter scheme is also known as the
replica-exchange mechanism. Of course, \D can also be compiled and run on a single node/CPU, in which case the only workgroup will be comprised 
of the only member -- the \textit{master} process.

\D does not provide a conventional domain decomposition mechanism which is commonly found in molecular dynamics packages 
such as \DP~ and Gromacs. Instead, \D employs the loop splitting approach which is in effect equivalent to \textit{particle decomposition}.
When combined with Verlet neighbour-lists, this decomposition approach becomes more efficient for large systems with cell dimensions 
significantly greater than the cut-off radius.

As is highlighted in Fig.~\ref{RE-workgroups}, the core loops in \D can be split at the level of either molecules or atoms \textit{within molecules} 
(but not both). This provides additional flexibility for optimization of parallel runs since, depending on the system topology, the user 
can choose between the two. Evidently, loop splitting becomes worthwhile only when the total number of iterations required for an entire loop 
(i.e. the total number of molecules or atoms in a molecule, denoted by \textbf{imN} and \textbf{iaM} in the figure), is considerably greater 
than the number of parallel processes to be employed. It is also important to keep each worker's loop completion time significantly 
longer than the time spent for inter-process communication per loop (as a guidance, \DD provides a comprehensive output of the communication 
times at the end of simulation). Therefore, optimisation with respect to the number of workers per workgroup is not a trivial task. In any 
case, it is obvious that one has to aim to parallelise (split) the loops that require a greater, rather than a smaller, number of iterations.

Particle decomposition (if invoked) can also be combined with an alternative approach, which is to run the simulation in the ``task farming'' 
mode. This mode is implemented in \DD as a special case of the replica exchange setup (the option invoked by the `use repexch' directive in 
the CONTROL input file) where the temperature increment between consecutive replicas is set to zero. In this case the configuration exchanges are 
omitted, and, thus, completely independent MC trajectories are generated for the same system, and the statistics accumulated by all the parallel
tasks can then be aggregated together.

\subsection{Performance improvements}
In developing version 2 of \DD we have made a number of changes to the implementation of algorithms in order to improve the performance of the 
code. These changes have primarily related to the reciprocal-space component of the Ewald summation. (However, 
performance improvements have also been realised for systems where the Ewald summation is not utilised).

Here we describe these improvements. 
In short, the first improvement was to alter the memory model in order to greatly reduce memory use, allowing multiple instances of \DD to be run
in parallel more efficiently; and the second improvement was to use a more efficient approach for calculating the reciprocal-space energy in 
orthorhombic systems. We elaborate on these improvements below.

\subsubsection{Ewald summation}
The classic approach to treating long-range interactions such as electrostatics is to use the Ewald summation~\cite{Allen_Tildesley_87}, which 
decomposes the interactions into real- and reciprocal-space components.
In MD the fact that the optimal scaling of the Ewald summation with system size $N$ is $\mathcal{O}(N^\frac{3}{2})$ has meant that alternatives have 
been sought. Accordingly methods of lower computational complexity, such as Smooth Particle Mesh Ewald (which has complexity $\mathcal{O}(N\ln(N))$),
have been developed and are used extensively.  In MC however we are typically moving a single atom or molecule at a time, and in this case 
the classic Ewald summation remains a competitive algorithm. Many excellent texts introduce, derive and discuss various approaches to 
the Ewald summation (e.g.~\cite{Allen_Tildesley_87,Frenkel_Smit}). Here we focus 
only on what is relevant to the improvements we have made in \Dns.

The reciprocal-space component of the electrostatics interactions, $U_{\text{rec}}$, is typically the most computationally expensive part of the 
Ewald summation calculation.  In practice this requires calculating a contribution from each charged particle in the system at a number of reciprocal
lattice vectors:
\begin{equation}
U_{\text{rec}}(\mathbf{r}_{1}, ...,\mathbf{r}_{N})=\frac{1}{V}\sum_{\mathbf{k\in\mathcal{S}}}\frac{4\pi}{|\mathbf{k}|^2}\exp(-|\mathbf{k}|^2/4\alpha^2)\bigg|\sum_{j}q_j\exp(i\mathbf{k}\cdot\mathbf{r}_j)\bigg|^2,
\label{eq:ULRdef}
\end{equation}
where $V$ is the volume of the system, $\mathbf{r}_i$ and $q_i$ are the position and charge of the $i$th atom, $N$ is the total number of atoms,
$\alpha$ is the Ewald parameter determining the relative ranges of the real- and reciprocal-space contributions, $i$ is the imaginary unit, 
$|\dotsc|$ denotes the complex modulus, and $\mathcal{S}$ is a set of reciprocal lattice vectors ($\mathbf{k}$ denotes a reciprocal lattice
vector) excluding $\mathbf{k}=0$.
Note that for a given set of vectors $\mathcal{S}$ (i.e. for a given simulation cell volume and shape) the coefficients preceding the modulus can be 
pre-computed and stored to improve efficiency. This is done in \Dns.

\subsubsection{Improved memory model}
$U_{\text{rec}}$ depends on atom $j$ through the quantities $\nu_{\mathbf{k}}\equiv q_j\exp(i\mathbf{k}\cdot\mathbf{r}_j)$ for all 
$\mathbf{k}\in\mathcal{S}$. The most expensive part of a computation of $U_{\text{rec}}$, or its change $\Delta U_{\text{rec}}$
when an atom is moved, is calculating these quantities. 
They can be calculated separately, meaning that when a given atom $j$ moves we only need to recalculate the 
set $\lbrace\nu_{\mathbf{k}}\rbrace$, and compare them with the old values of $\lbrace\nu_{\mathbf{k}}\rbrace$, in order to obtain 
$\Delta U_{\text{rec}}$. Their `new values' of course must be calculated explicitly. 
However there is a choice as to whether to store their `old' values in memory during the simulation every move, or calculate them afresh every move. 
The former approach costs memory, but in theory is more efficient because it eliminates the need to recalculate the old values every move.

In \D version 1 the old $\lbrace\nu_{j,\mathbf{k}}\rbrace$ were stored (meaning that the old values of $\lbrace\nu_{\mathbf{k}}\rbrace$ did not 
have to be recalculated every move). However for system sizes where $N\ge10^3$ the 
memory requirements become large.  Additionally we found in \DD version 2 that for systems of this size the memory access times 
associated with this approach typically led to run times which are slower than if both old and new $\lbrace\nu_{\mathbf{k}}\rbrace$
were calculated afresh every move.
For modern multi-core processors the reduced memory overhead of the latter approach also improved the \emph{trivial parallelism} performance 
when many instances of \DD were run on a single processor. The improved performance, associated with the reduced memory overhead, comes despite
the doubling in the processing cost associated with recalculating the old $\lbrace\nu_{j,\mathbf{k}}\rbrace$ every move.

\subsubsection{Improved implementation for orthorhombic systems}
In an orthorhombic simulation cell the lattice vectors of the system are orthogonal, and the same applies to its reciprocal lattice vectors. In this 
case $\exp(i\mathbf{k}\cdot \mathbf{r}_{j})$ can be decomposed as follows:
\begin{equation}
\exp(i\mathbf{k}\cdot \mathbf{r}_{j})=\exp(ik_{x}r_{j,x})\exp(ik_{y}r_{j,y})\exp(ik_{z}r_{j,z}),
\label{eq:expdecomp}
\end{equation}
where $k_{x}$ and $r_{x}$ are the $x$ components of $\mathbf{k}$ and $\mathbf{r}$, respectively, and similarly for the
$y$ and $z$ components. Note that $\mathbf{k}$ is a linear combination of the three primitive reciprocal lattice vectors 
$(k_{0,x},0,0)$, $(0,k_{0,y},0)$ and $(0,0,k_{0,z})$. In other words $k_x=n_{x}k_{0,x}$, where $n_x$ is an integer, and similarly for $k_y$ and 
$k_z$. Hence $\exp(i\mathbf{k}\cdot \mathbf{r}_{j})$ can be expressed as 
\begin{equation}
\exp(i\mathbf{k}\cdot \mathbf{r}_{j})=\exp(ik_{0,x}r_{j,x})^{n_x}\exp(ik_{0,y}r_{j,y})^{n_y}\exp(ik_{0,z}r_{j,z})^{n_z}.
\label{eq:expdecomp2}
\end{equation}
This implies that $\exp(i\mathbf{k}\cdot \mathbf{r}_{j})$ could be obtained by first calculating $\exp(ik_{0,x}r_{j,x})$, 
$\exp(ik_{0,y}r_{j,y})$ and $\exp(ik_{0,z}r_{j,z})$, storing these three (complex) values, and then using them to obtain 
$\exp(i\mathbf{k}\cdot \mathbf{r}_{j})$ via multiplication according to the above equation.
This would involve only three calls to the exponential function, and is significantly more efficient than evaluating 
$\exp(i\mathbf{k}\cdot \mathbf{r}_{j})$ afresh for each $\mathbf{k}$, which involves one call to the exponential function 
per $\mathbf{k}$ vector.

The above approach can be generalised to systems with non-orthogonal lattice vectors. However we have yet to implement the general approach
in \Dns -- only orthorhombic systems are currently supported (see below) -- since doing so would involve largescale refactoring throughout \Dns.

\subsubsection{Quantifying the improvements}
\begin{figure}
\begin{center}
\includegraphics[width=0.7\textwidth,angle=270]{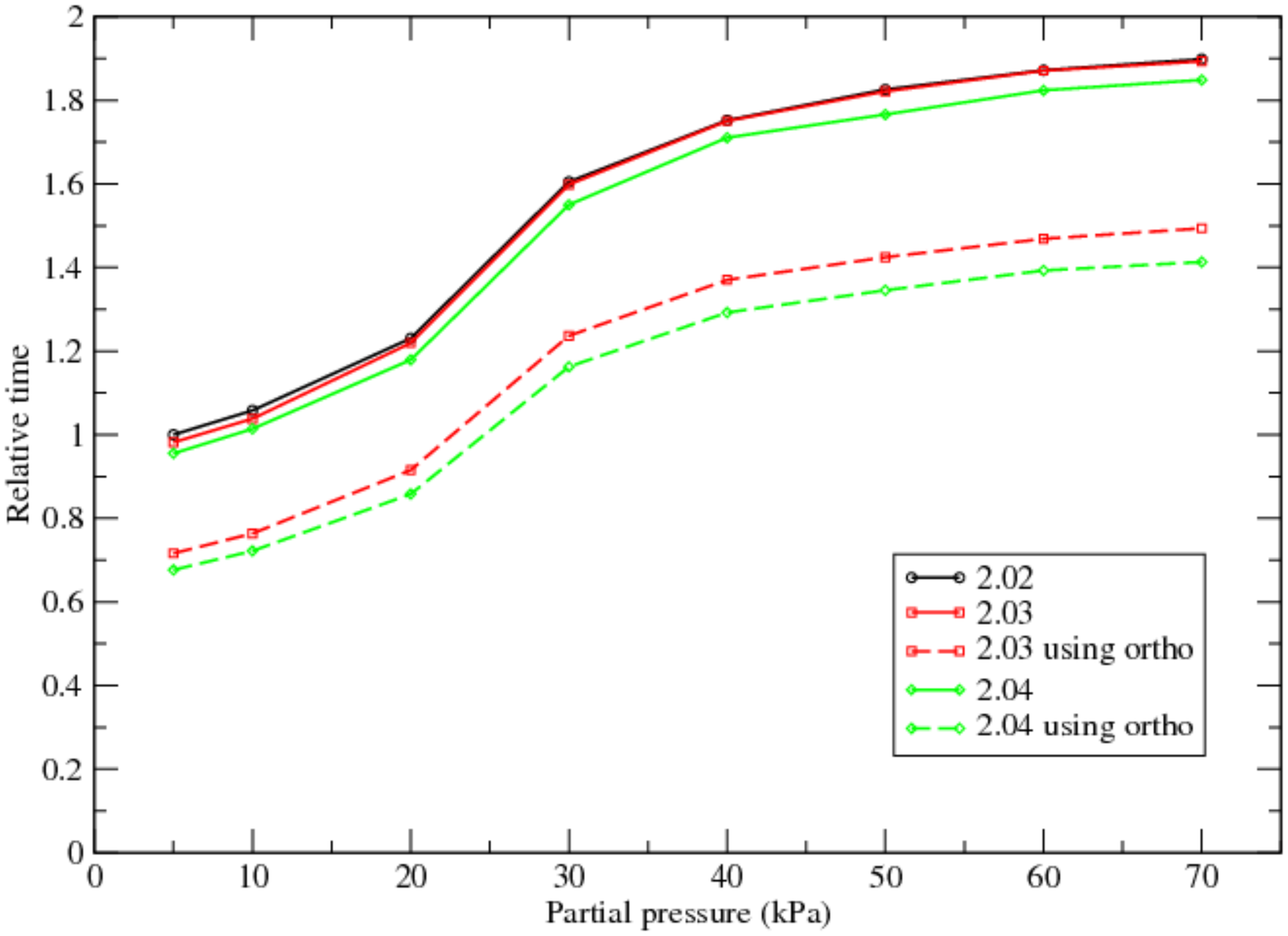}
\caption{Performance of \D for the adsorption isotherm of $\mathrm{CO}_2$ in the metal organic framework IRMOF-1 at 208~K given in~\cite{Gowers_2018},
with and without the `use ortho' directive (which enables the improvements described in Appendix~\ref{sec:performance}). Data is included
for three recent versions of \Dns, as labeled in the legend.}
\label{fig_gowers_test}
\end{center}
\end{figure}

The above improvements are available in \Dns for orthorhombic simulation cells. However, they are not currently used by default, and need to
be enabled by invoking the `use ortho' directive in the CONTROL file.
To illustrate the improvements, we ran the recently published test suite~\cite{Gowers_2018} which traces an adsorption isotherm of $\mathrm{CO}_2$ in 
the metal organic framework IRMOF-1 at 208~K. Simulations of 100,000 moves were run for the 8 partial pressures in the test suite,
with two instances run at each pressure in order to fully populate a 16 code node. Performance is illustrated in Fig.~\ref{fig_gowers_test} 
for three recent releases of \D compiled with the Intel compiler, with and without the use of the `use ortho' directive. Note that in GCMC
simulations \D does not use neighbour lists. This, combined with the large cut-off length employed in test system (25\AA), results in the 
computation time being dominated by the short range interactions. However we still see performance improvements of up to $\frac{1}{3}$.

\section{Example Python script utilising the toolkit}\label{sec:pythonscript}
The Python script below runs \D simulations and calculates the mean energy 
at various temperatures, as described in Section~\ref{sec:toolkit_interface}.

\begin{verbatim}
import os
import htk.sources.dlmonte as dlmonte
# List of temperatures to perform simulations at    
temperatures = [300, 310, 320, 330]
# Import input parameters from directory 'input' into a DLMonteInput object
dlminput = dlmonte.DLMonteInput.from_directory("input")
# Set up a DLMonteRunner for executing DL_MONTE from within Python
dlmrunner = dlmonte.DLMonteRunner("/bin/DLMONTE-SRL.X")
for T in temperatures:
    # Create the directory corresponding to this temperature 
    simdir = str(T)
    os.mkdir(simdir)
    # Amend the temperature in the input
    dlminput.control.main_block.statements["temperature"] = T
    # Create relevant input files in the simulation directory   
    dlminput.to_directory(simdir)
    # Run the simulation in that directory  
    dlmrunner.directory = simdir
    dlmrunner.execute()
    # Import the data generated by the simulation into a DLMonteOutput object
    dlmdata = dlmonte.DLMonteOutput.load(simdir)
    # Extract the energy timeseries from the data        
    energies = dlmdata.yamldata.time_series("energy")
    # Print the temperature and the corresponding average energy    
    print T, sum(energies)/len(energies)
\end{verbatim}

%
%
%



\end{document}